\documentclass{optica-article}

\journal{opticajournal} 

\articletype{Research Article}

\usepackage{todonotes}
\usepackage{physics}
\usepackage[labelformat=simple]{subcaption} 
     
\usepackage{float}          
\usepackage{booktabs}       
\usepackage{multirow}       
\usepackage[none]{hyphenat} 
\usepackage{siunitx}        
\usepackage{standalone}     
\newcommand{\ra}[1]{\renewcommand{\arraystretch}{#1}}
\usepackage{xparse}

\newcommand{\vep}{\ensuremath{\varepsilon}}

\makeatletter
\renewcommand*\env@matrix[1][\arraystretch]{%
  \edef\arraystretch{#1}%
  \hskip -\arraycolsep
  \let\@ifnextchar\new@ifnextchar
  \array{*\c@MaxMatrixCols c}}
\makeatother

\usepackage{array}
\newcommand{\PreserveBackslash}[1]{\let\temp=\\#1\let\\=\temp}
\newcolumntype{C}[1]{>{\PreserveBackslash\centering}p{#1}}
\newcolumntype{R}[1]{>{\PreserveBackslash\raggedleft}p{#1}}
\newcolumntype{L}[1]{>{\PreserveBackslash\raggedright}p{#1}}

\def\BibTeX{{\rm B\kern-.05em{\sc i\kern-.025em b}\kern-.08em
    T\kern-.1667em\lower.7ex\hbox{E}\kern-.125emX}}

\begin{document}

\title{Inverse Design for Waveguide Dispersion with a Differentiable Mode Solver}

\author{
    Dodd Gray,\authormark{1,2,*}
    Gavin N. West,\authormark{1}
    Rajeev J. Ram\authormark{1}
}

\address{
    \authormark{1}MIT, 77 Massachusetts Avenue, Cambridge, MA 02139, USA \\
    \authormark{2}MIT Lincoln Laboratory, 244 Wood Street, Lexington, MA 02421, USA
}

\email{\authormark{*}dodd.gray@ll.mit.edu}

\begin{abstract*} 
Inverse design of optical components based on adjoint sensitivity analysis has the potential to address the most challenging photonic engineering problems.
However existing inverse design tools based on finite-difference-time-domain (FDTD) models are poorly suited for optimizing waveguide modes for adiabatic transformation or perturbative coupling, which lies at the heart of many important photonic devices.
Among these, dispersion engineering of optical waveguides is especially challenging in ultrafast and nonlinear optical applications involving broad optical bandwidths and frequency-dependent anisotropic dielectric material response. 
In this work we develop gradient back-propagation through a general purpose electromagnetic eigenmode solver and use it to demonstrate waveguide dispersion optimization for second harmonic generation with maximized phase-matching bandwidth.
This optimization of three design parameters converges in eight steps, reducing the computational cost of optimization by $\sim$100x compared to exhaustive search and identifies new designs for broadband optical frequency doubling of laser sources in the 1.3--\SI{1.4}{\um} wavelength range. 
Furthermore we demonstrate that the computational cost of gradient back-propagation is independent of the number of parameters, as required for optimization of complex geometries.
This technique enables practical inverse design for a broad range of previously intractable photonic devices.  
\end{abstract*}

\section{Introduction}
Inverse design methods based on adjoint sensitivity analysis have been demonstrated as a promising toolkit for photonic design \cite{NatRev}, enabling tunable high performance devices through optimization of complex designs with large numbers of parameters.
These techniques generally rely on gradients of electromagnetic models calculated using the adjoint method with a computational cost that is independent of the number of optimized parameters.
A variety of mature software tools exist for inverse design of optical devices based on differentiable finite-difference-time-domain (FDTD) \cite{OskooiMEEP,Hammond2022High} and finite-difference-frequency-domain (FDFD) models \cite{Piggott2017Fabrication}.
These have been used to optimize light scattering between fixed input and output modes \cite{PiggottDichro,MichaelsGC,ChungLens,ChristiansenNP} for a wide range of applications but are poorly suited for optimization of the modal dispersion of optical resonances and waveguide modes.

Modeling of waveguide mode fields and modal dispersion is ubiquitous and indispensable in integrated photonics. 
Many important design problems, especially those involving adiabatic or perturbative coupling of waveguide modes require differentiable mode solving for compatibility with sensitivity analysis and inverse design.
The efficiency of nonlinear optical interactions are particularly sensitive to dispersion \cite{JankUB,NonlinearCavityPairProduction,tflnOPA} because coupling is achieved over large optical bandwidths and long propagation lengths.
Various approaches to modal dispersion optimization have been demonstrated, including neural networks \cite{ma2021deep,alagappan2023NN,el2014NN,malheiros2012NN}, model order reduction \cite{talentiKdotP}, automatic differentiation \cite{MinkovPhC,nasidiPhC,vercruyssePhC,nussbaum2021PhC,vial2022autodiff}, and others \cite{JohnsonSDP,CastelloSiWG,hameedTrustRegion,stainko2007TO}.
None of these simultaneously support arbitrary, non-periodic waveguide cross sections and account for dielectric material anisotropy and frequency-dependence, limiting their general utility for photonic design.

Surprisingly, optimization of dispersion-related objective functions with only two or three geometry parameters can be computational expensive.
For example second harmonic generation (SHG) or three-wave mixing involve frequencies that span at least one octave.
This wide optical bandwidth, coupled with material birefringence often requires a large number of modes to be modeled (>10) for each set of geometric parameters and wavelengths.
The underlying material anisotropy also necessitates solving the Helmholtz Equation with tensor-valued dielectric constants and introduces mode degeneracies which can frustrate numerical convergence.
Finally, the broadband nonlinear phenomena considered here are extremely sensitive to small changes in geometry – a fact that has driven the development of adaptive fabrication strategies in thin film lithium niobate (TFLN) on insulator waveguides \cite{xin2024wavelength,chen2024adapted} – necessitating the use of a fine discretization mesh and a large number of simulations for brute-force optimization of the geometry\footnote[1]{
    We consider the design problem discussed in Section 3. On a conventional workstation (16 cores, 3 GHz clock speed, 32 GB RAM) robustly solving the Helmholtz Equation for fundamental and second harmonic TE$_{00}$ modes in isotropic media would require \SI{10}{\s} (4 modes), however the corresponding simulations with anisotropic materials require \SI{60}{\s} (12 modes).
    A coarse sweep of waveguide geometry parameters (20 widths $\times$ 20 thicknesses $\times$ 10 etch depths) considering SHG of a single wavelength requires 67 hours (48000 modes) if solved in serial, or 8 hours with 8 mode solvers running concurrently.
    This increases linearly with wavelengths and exponentially with parameters.
    }.
Indeed, the existing literature on SHG bandwidth optimization only considers two \cite{JankUB} or three \cite{fergestad2020ultrabroadband,zhang2022ultrabroadband} waveguide geometry parameters (eg. width, thickness, and etch depth).
Additional design parameters are easily accessible experimentally (eg. upper and lower cladding thickness and side-wall angle) but have not been considered because of the exponential scaling of computational effort.
Clearly a more efficient approach to optimization is required.
Here, differentiable mode solvers with gradient-based optimization are shown to make such dispersion-related optimization problems tractable.
This class of inverse design tools efficiently manages the sensitivity to geometry by using parameter gradients to adaptively step through geometric parameters even in the presence of strong material dispersion and anisotropy.
As we demonstrate below, these gradients are calculated efficiently for a large number of parameters by exploiting automatic differentiation.
Our implementation enables incorporation of waveguide mode solutions as a modular component in larger differentiable models, including concurrent evaluation of mode solutions at multiple frequencies for time-efficient broadband dispersion optimization.

In Section~\ref{sec:implementation} we describe the implementation of gradient backpropagation through a fully vectorial plane-wave expansion mode solver compatible with anisotropic and frequency-dependent dielectric materials.
In Section~\ref{sec:demonstration} we demonstrate its application to waveguide dispersion optimization for nonlinear optical applications. In Section~\ref{sec:performance} we describe the performance and scaling of our differentiable mode solver. Finally in Section~\ref{sec:conclusion} we summarize our results and discuss future work.

\section{Implementation}\label{sec:implementation}

\begin{figure}[htbp]
    \centering{
    \phantomsubcaption\label{fig:algorithm:a}
    \phantomsubcaption\label{fig:algorithm:b} 
    \phantomsubcaption\label{fig:algorithm:c}
    \phantomsubcaption\label{fig:algorithm:d}
    \phantomsubcaption\label{fig:algorithm:e} 
    \phantomsubcaption\label{fig:algorithm:f}
    \phantomsubcaption\label{fig:algorithm:g}
    \phantomsubcaption\label{fig:algorithm:h} 
    \phantomsubcaption\label{fig:algorithm:i}
    \phantomsubcaption\label{fig:algorithm:j}
    \phantomsubcaption\label{fig:algorithm:k}
    }    
    \includegraphics[width=\textwidth]{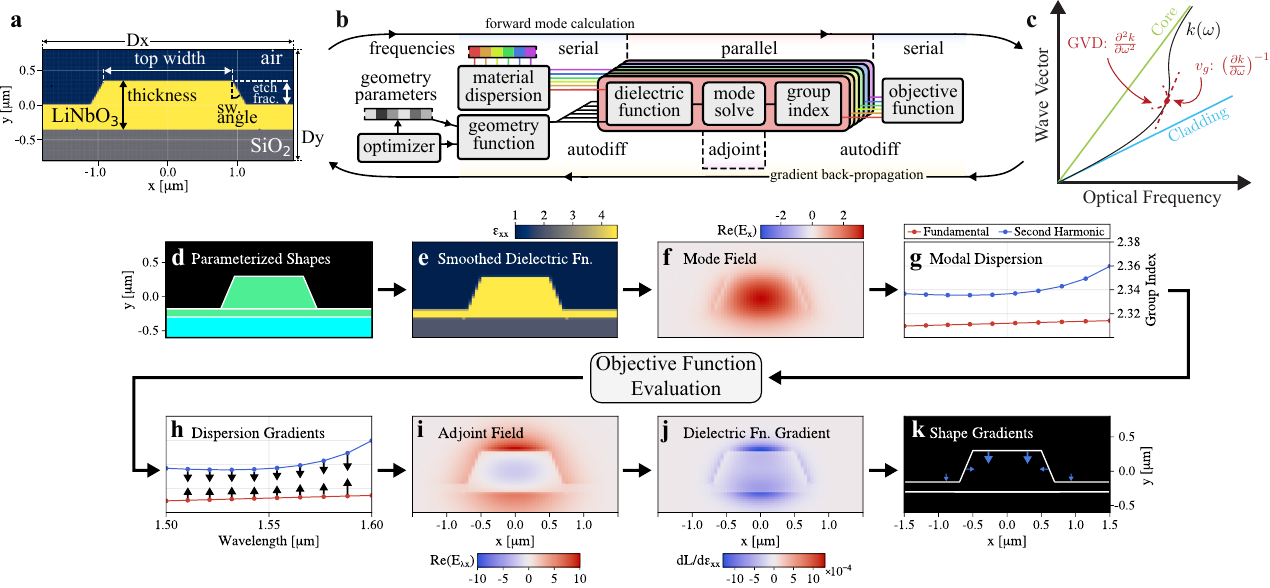}
    \caption{
    \subref{fig:algorithm:a} Rib waveguide design parameterized by width, thickness and etch depth. 
    \subref{fig:algorithm:b} Block diagram showing computational modeling of waveguide modes and a cost function dependent on their 
    \subref{fig:algorithm:c} dispersion, then back-propagation of cost function gradient components to the input design parameters.  
    One optimization step consists of the following steps:
    \subref{fig:algorithm:d} instantiate the parametric shapes and assign bulk materials. 
    Next
    \subref{fig:algorithm:e} generate a smooth dielectric function.
    \subref{fig:algorithm:f} Eigenmodes are solved using a plane-wave expansion-based solver. 
    The modal indices and field distributions are used to calculate
    \subref{fig:algorithm:g} the modal dispersion (here the effective group index) of the specified waveguide  
    In the final step of the forward calculation we compute a dispersion and/or mode-field dependent objective function quantifying waveguide performance.   
    Gradients then propagate backwards from the objective function via automatic differentiation to
    \subref{fig:algorithm:h} gradients with respect to the computed waveguide modes and their dispersion. 
    To back-propagate gradient components through mode solutions we compute
    \subref{fig:algorithm:i} corresponding adjoint fields for each mode impacting the objective function. 
    From the adjoint fields we calculate
    \subref{fig:algorithm:j} Gradients with respect to the smoothed dielectric tensor at each grid point. 
    Finally we use automatic differentiation to calculate
    \subref{fig:algorithm:k} gradients with respect to the input design parameters, which are shape parameters in this example.
    The initial parameters, objective function value, and parameter gradients are passed to a gradient-based optimizer which computes updated parameter values for the next iteration. 
    }
    \label{fig:algorithm}   
\end{figure}

The workflow for inverse design of waveguides with optimized modal dispersion is depicted in Fig.~\ref{fig:algorithm:a}\textendash\ref{fig:algorithm:c}.
First a function is defined mapping input design parameters to dielectric tensor values on a discretized spatial grid.
An example of such parametric waveguide design is shown in Fig.~\ref{fig:algorithm:a}. 
The computational steps we used to numerically model and optimize waveguide modes for a desired dispersion are schematically depicted in Fig.~\ref{fig:algorithm:b}.
We use desired modal properties illustrated in Fig.~\ref{fig:algorithm:c} such as effective index, group velocity (or group index) and group velocity dispersion (GVD) at target optical frequencies to determine optimal waveguide designs.
To achieve this we supplied with the value and sensitivity to design parameters of an objective function quantifying a desired dispersion property to a generic optimization algorithm \cite{nlopt}.
Allowing the optimization algorithm to iteratively update design parameters in a feedback loop rapidly improved the cost function, typically converging on a local optimum after only a handful of iterations.
The sequence of calculations comprising a single iteration is shown in Fig.~\ref{fig:algorithm:d}\textendash\ref{fig:algorithm:k} and described below.

\subsection{Parametric Geometry and Materials}
To model resonant modes of a dielectric waveguide we write a function mapping a set of shape and material parameters to dielectric tensor values at each spatial grid point in our simulation. 
Figure~\ref{fig:algorithm:d} illustrates a geometry instantiated through a constructive geometry library \cite{GeomPrims} which provides parametric shape primitives to programatically generate waveguide geometries as lists of shapes with assigned bulk materials.

We implement an interface for arbitrary models of material dielectric tensors as a function of frequency and other parameters, such as temperature, alloy concentration, and orientation. 
The first and second derivatives w.r.t frequency of all material bulk dielectric tensors, $d\vep/d\omega$ and $d^2\vep/d\omega^2$, are calculated using a computer algebra package \cite{gowda2021high}.
These derivatives are later used in the calculation of group velocity and group velocity dispersion. 
Analytic pre-computation is inexpensive, improves numerical precision, and reduces the burden on later auto-differentiation steps.

\subsection{Dielectric Smoothing}
Once a geometry is specified and material dielectric tensors are computed, the dielectric distribution must be smoothed across geometric/material boundaries, seen in Fig.~\ref{fig:algorithm:e}. 
This ensures that the resonant frequencies and mode fields of electromagnetic eigenmode solutions vary smoothly with design parameters \cite{OskooiSmoothing}, and corrects for spatial discretization errors to first order.
We implement the anisotropic dielectric smoothing algorithm developed by Kottke, et al. \cite{KottkeSPS} to calculate values of the smoothed dielectric tensor, here denoted $\tilde{\vep}$, and extend the algorithm to calculate first and second frequency derivatives of the smoothed tensor. 
Near a dielectric interface with surface normal $\hat{n}$ between materials with dielectric tensors $\vep_1$ and $\vep_2$, 

\begin{equation}
    \tilde{\vep} = \mathbf{R}_{\hat{n}} \left[ \tau^{-1}\left( r\tau( \mathbf{R}_{\hat{n}}^T \vep_1 \mathbf{R}_{\hat{n}} ) + (1-r)\tau( \mathbf{R}_{\hat{n}}^T \vep_2 \mathbf{R}_{\hat{n}} ) \right) \right] \mathbf{R}_{\hat{n}}^T \label{eq:KottkeSmoothing1} 
\end{equation}

where $\mathbf{R}_{\hat{n}}$ is a rotation matrix mapping between the grid axes and a coordinate frame with the first coordinate along the local surface normal of the material interface $\hat{n}$, $r$ is the fill fraction of material 1 in the grid pixel or voxel, and $\tau(\vep)$ and $\tau^{-1}(\tau)$ are 3$\times$3 matrix-valued functions of 3$\times$3 matrices derived in \cite{KottkeSPS}.
The operator $\mathbf{R}_{\hat{n}}$ is constructed as a function of $\hat{n}$ by ortho-normalization of successive cross products

\begin{equation}
    \mathbf{R}_{\hat{n}}(\hat{n}) = \begin{bmatrix} \hat{n}  & \frac{\hat{n} \times \hat{z}}{| \hat{n} \times \hat{z} |} & \frac{ \hat{n} \times \frac{\hat{n} \times \hat{z}}{| \hat{n} \times \hat{z} |}  }{ \left| \hat{n} \times \frac{\hat{n} \times \hat{z}}{| \hat{n} \times \hat{z} |} \right| } \end{bmatrix} \label{eq:orthonormalization}
\end{equation}

which is a differentiable function of $\hat{n}$ as long as $\hat{n}$ is not parallel to $\hat{z}$. For 3D models a generalized form Eq.~\ref{eq:orthonormalization} can be used to be continuous for all $\hat{n}$ \cite{Zhou_2019_CVPR}.
Equations~\ref{eq:KottkeSmoothing1}-\ref{eq:orthonormalization} define a differentiable map from $r$, $\hat{n}$, $\vep_1$ and $\vep_2$ to $\tilde{\vep}$, 

\begin{equation}
    \mathrm{vec}(\tilde{\vep}) = F_{\tilde{\vep}}(\mathrm{vec}(r,\hat{n},\vep_1,\vep_2)) \label{eq:FlattenedSmoothingFn}
\end{equation}

where $\mathrm{vec}(\dots)$ denotes flattening into a vector. We use a computer algebra system \cite{gowda2021high} to analytically compute and pre-compile functions for the Jacobian and Hessian matrices $J_{\tilde{\vep}}$ and $H_{\tilde{\vep}}$ of the function $F_{\tilde{\vep}}$ in Eqn.~\ref{eq:FlattenedSmoothingFn}.
With these we calculate the first two frequency derivatives of the smoothed dielectric tensor $d\tilde{\vep}/d\omega$ and $d^2\tilde{\vep}/d\omega^2$ at each spatial grid point as functions of the bulk dielectric tensors $\vep_1$ and $\vep_2$ and their first two frequency derivatives

\begin{align}
    d\tilde{\vep}/d\omega =& J_{\tilde{\vep}} \vec{v}_1 \label{eq:dEpsdOmega} \\
    d^2\tilde{\vep}/d\omega^2 =& \vec{v}_1^T H_{\tilde{\vep}} \vec{v}_1 + J_{\tilde{\vep}} \vec{v}_2 \label{eq:dsqEpsdOmegasq}
\end{align}

where $\vec{v}_1$ and $\vec{v}_2$ are vectors containing the first and second derivatives of the material dielectric tensors w.r.t frequency and zeros corresponding to $r$ and $\hat{n}$ which are frequency independent,

\begin{align}
    \vec{v}_1 =& \mathrm{vec}( 0, [0,0,0], d\vep_1/d\omega, d\vep_2/d\omega) ) \label{eq:v1} \\
    \vec{v}_2 =& \mathrm{vec}( 0, [0,0,0], d^2\vep_1/d\omega^2, d^2\vep_2/d\omega^2) ) \label{eq:v2}
\end{align}

Moving forward we will refer to the smoothed dielectric tensor simply as $\vep$ to simplify notation. 

\subsection{Electromagnetic Eigenmode Solver}
Our differentiable electromagnetic mode solver is adapted from the plane-wave expansion algorithm in the open-source MIT Photonic Bands (MPB) \cite{JohnsonMPB} software package.
The algorithm leverages discrete Fourier transforms (DFTs) \cite{FrigoFFTW} to efficiently apply curl and dielectric tensor operations to the electromagnetic mode field in reciprocal and real space, respectively.
We used the Julia programming language \cite{julialang} to implement a matrix-free discretized Helmholtz operator and used generic iterative solvers based on Krylov subspace methods \cite{IterativeSolvers,krylovJL,DFTKjcon} for partial eigen-decomposition to solve for a specified number of approximate eigenvector-eigenvalue pairs.
Our primary motivation for using Julia was compatibility with automatic differentiation, discussed below in Sec.~\ref{subsec:AD}.
In this work we focus on laterally confined waveguide modes, but the following analysis does not require this assumption.
For reasons discussed below it is most convenient to solve Eq.~\ref{eq:HelmholtzOp} with periodic boundary conditions, despite the fact that we are interested in dielectric waveguide geometries which are not periodic in the dimensions transverse to the propagation axis (here the $x-y$ plane).

We consider a Helmholtz Equation for the magnetic field $H_{\rm n}$ of an electromagnetic waveguide mode propagating in the $+\hat{z}$ direction (along the waveguide axis), parameterized by dielectric tensor values $\vep(x,y,z)$ at each point on a regular spatial grid $ \left\{ x_{\rm i},y_{\rm i},z_{\rm i} \right\} $ and a wave vector $k\hat{z}$ with magnitude $k$

\begin{equation}
    \mathbf{M}(k,\vep) H_{\rm n} = \omega_{\rm n}^2 H_{\rm n} \label{eq:HelmholtzEq}
\end{equation}

The Helmholtz operator $\mathbf{M}$ consists of two curl operations sandwiching multiplication by the inverse dielectric tensor $\vep^{-1}$. 

\begin{equation}
    \mathbf{M}(k,\vep)  = \left[\nabla \times\right]  \left[\vep^{-1} \cdot \right]  \left[\nabla \times\right] \label{eq:HelmholtzOp}
\end{equation}

Expressing the mode field and curl operators in a plane wave basis we have

\begin{equation}
    \mathbf{M}(k,\vep) \Tilde{h}_{\rm n}  = \left[ \mathbf{C}^\dagger \vep^{-1}  \mathbf{C} \right] \Tilde{h}_{\rm n} = \omega^2_{\rm n} \Tilde{h}_{\rm n} \label{eq:PWHelmholtzEq}
\end{equation}

where $\Tilde{h}_{\rm n}$ is the projection of $H_{\rm n}$ onto transversely polarized plane wave components at each reciprocal lattice point and $\mathbf{C}$ represents the fused DFT, Cartesian-to-transverse projection, and curl operators.
The electromagnetic Helmholtz equation as expressed in Eq.~\ref{eq:PWHelmholtzEq} enables computationally efficient approximate numerical eigenmode solutions \cite{JohnsonMPB}.
In many dispersion engineering applications it is important to account for the dependence of the dielectric tensor $\vep$ on optical frequency $\omega$. 
In this case the Helmholtz equation becomes a nonlinear eigenproblem with an eigenvalue-dependent operator.

    \begin{equation}
        \mathbf{M}\left(k,\varepsilon(\omega_{\rm n}) \right) H_{\rm n} = \omega_{\rm n}^2 H_{\rm n}. \label{eq:NonlinearHelmholtzEqn}
    \end{equation}

Solutions to Eq.~\ref{eq:NonlinearHelmholtzEqn} are found by iteratively solving Eq.~\ref{eq:PWHelmholtzEq} using a Newton method to update $k$. 
We will refer to Eq.~\ref{eq:PWHelmholtzEq} as the ``fixed-$k$'' eigenproblem and Eq.~\ref{eq:NonlinearHelmholtzEqn} as the ``fixed-$\omega$'' eigenproblem.
Supporting derivations and the definition of $H_{\rm n}$, $\Tilde{h}_{\rm n}$, and $\mathbf{C}$ are available in Supplement 1.
For mode sorting and subsequent calculations, we transform $\Tilde{h}_{\rm n}$ to a real-space electric field mode $E_n$ as shown in Fig.~\ref{fig:algorithm:f}.

\subsection{Modal Group Velocities, GVDs and the Objective Function}

To calculate frequency dispersion relationships such as shown in Fig.~\ref{fig:algorithm:g} we derive AD-compatible expressions for the frequency derivatives of the modal index. 
The modal group index is calculated as the ratio between Poynting flux and energy density in mode volume \cite{jacksonEM}.
We express the modal group index in terms of the plane-wave basis fields as 

    \begin{equation}
        n_{\rm g,n} = \frac{ \omega_{\rm n}^2 + \frac{1}{2}\omega_{\rm n} \sum \Tilde{h}_{\rm n}^\dagger \left[ \mathbf{C}^\dagger \vep^{-1} \frac{d\vep}{d\omega} \vep^{-1} \mathbf{C} \right] \Tilde{h}_{\rm n} }{ \omega_{\rm n}\sum \Tilde{h}_{\rm n}^\dagger \frac{\partial \mathbf{M}}{\partial k} \Tilde{h}_{\rm n} } \label{eq:pw_group_index}
    \end{equation}

where $\frac{d \vep}{d \omega}$ is the first frequency derivative of the smoothed dielectric tensor defined in Eq.~\ref{eq:dEpsdOmega}.

The modal group velocity dispersion (GVD) $\text{GVD}_n = \frac{\text{d} n_{g,n}}{\text{d}\omega}$ can also be computed from a single mode solution using the adjoint field corresponding to group index and the second frequency derivative of the smoothed dielectric tensor, $\frac{d^2 \vep}{d \omega^2}$, defined in Eq.~\ref{eq:dsqEpsdOmegasq}.
Expressions for the GVD are derived in Supplement 1. Group index and GVD values calculated directly from a single mode solution avoid numerical error from finite difference approximation \cite{nocedal2006} and can be optimized using fewer total eigen- and adjoint problem solutions.

In the final step of the forward calculation we compute an objective function quantifying the waveguide performance. The objective function can depend in an arbitrary way on the calculated mode fields and dispersion quantities as well as the input parameters and intermediate quantities such as the dielectric function.  

\subsection{Automatic Differentiation}\label{subsec:AD}
Efficient calculation of the objective function value and gradient is highly desirable when optimizing of objective functions in high-dimensional parameter spaces \cite{nocedal2006}.
The standard Julia automatic differentiation framework provides extensive compatibility with native Julia functions \cite{ChainRulesJL}, enabling convenient compiled gradient computations of arbitrary user-defined functions.

We defined ``pullback'' functions for the fixed-$k$ and fixed-$\omega$ Helmholtz eigen-problems using the adjoint method to map partial derivatives with respect to mode fields and wave vector magnitudes back to partial derivatives w.r.t the optical frequency, dielectric tensor elements, and material dispersion at each spatial grid point (the Helmholtz operator parameters), described in Sec.~\ref{sec:adjoint_analysis}.
These custom AD rules enable differentiation of cost functions with arbitrary dependence on mode solutions.
They do not require adjustment or re-definition when the mode simulation or cost function is modified. 
We note that an AD-compatible constructive geometry package is available in Julia \cite{GeomPrims}, enabling propagation of gradients w.r.t the dielectric function (Fig.~\ref{fig:algorithm:j}) to gradients in the shape parameters (Fig.~\ref{fig:algorithm:k}). 

\subsection{Adjoint Sensitivity Analysis}\label{sec:adjoint_analysis}

To provide an unbroken chain of derivatives to map sensitivities between the objective function and input parameters, we apply the adjoint method to solutions of Eq.~\ref{eq:NonlinearHelmholtzEqn}.
We begin by following the approach of \cite{johnsonAdj} to derive the adjoint problem for the fixed-$k$ Helmholtz equation, then extend this derivation to the fixed-$\omega$ Helmholtz equation.
The sensitivities of the objective function with respect to the wave vector magnitude $k$ and smoothed permittivity $\vep$ for the linear problem are given by

    \begin{align}
        \frac{\partial g}{\partial k} = & \frac{\partial g}{\partial \omega_{\rm n}^2} \frac{\partial \omega^2_{\rm n}}{\partial k} + \sum_j \Tr\left( \mqty[ -(v_{j\rm z}\hat{n}_j + n_{j\rm z} \vec{v}_j)^T  \\  (v_{j\rm z}\hat{m}_j + m_{j\rm z} \vec{v}_j)^T ] \left[ \Tilde{E}_{\rm n} \Tilde{h}_{\lambda \rm n}^\dag + \Tilde{E}_{\rm \lambda n} \Tilde{h}_{\rm n}^\dag \right]\right) \label{eq:KGrad} \\
        \frac{\partial g}{\partial \vep} =&  E_{\rm \lambda n} E_{\rm n}^\dag \label{eq:EpsilonGrad} 
    \end{align}

\noindent where $\Tilde{h}_{\lambda \rm n}$ is the adjoint mode field shown in Fig.~\ref{fig:algorithm:h} and where we have used the simplifying transformations

    \begin{align*}
        \frac{\partial\omega_n^2}{\partial k} = &  2 \frac{\omega_n}{n_{g,n}} \\ 
        \vec{v}_j = & k\hat{z} + \vec{g}_j
    \end{align*}

\noindent where each $\vec{g}_{\rm j}$ is a reciprocal lattice point in the plane wave basis. We introduce the nonlinearity of Eq.~\ref{eq:NonlinearHelmholtzEqn} through an equality constraint (see Supplement 1) and differentiate to find

    \begin{align}
        \frac{\partial k}{\partial \vep} = &  -\frac{\partial k}{\partial \omega^2} E_{\rm n} E_{\rm n}^\dag \label{eq:EqDkDepsNL} \\
        \frac{\partial g}{\partial \omega^2} =& \left[ \frac{\partial g}{\partial k}  + \frac{\partial g}{\partial \tilde{h}_{\rm n}} \frac{\partial \tilde{h}_{\rm n}}{\partial k}\right] \frac{\partial k}{\partial \omega^2} \label{eq:EqDgDomsqNL} \\
        \frac{\partial g}{\partial \vep} =&  \frac{\partial g}{\partial \omega^2} \frac{\partial \omega^2}{\partial \vep}  \label{eq:EqDgDepsNL}
    \end{align}

\noindent where in Eq.~\ref{eq:EqDgDomsqNL} 

    \begin{equation}
        \frac{\partial g}{\partial \tilde{h}_{\rm n}} \frac{\partial \tilde{h}_{\rm n}}{\partial k} = \sum_j \Tr\left( \mqty[ -(v_{j\rm z}\hat{n}_j + n_{j\rm z} \vec{v}_j)^T  \\  (v_{j\rm z}\hat{m}_j + m_{j\rm z} \vec{v}_j)^T ] \left[ \Tilde{E}_{\rm n} \Tilde{h}_{\lambda \rm n}^\dag + \Tilde{E}_{\rm \lambda n} \Tilde{h}_{\rm n}^\dag \right]\right) \label{eq:dh_dk_clarification}
    \end{equation}

\noindent as in Eq.~\ref{eq:KGrad} and requires calculation of the adjoint field. Using Eqs.~\ref{eq:EqDkDepsNL}-\ref{eq:EqDgDepsNL} we define a $\left\{ \frac{\partial g}{\partial k}, \frac{\partial g}{\partial \Tilde{h}_{\rm n}} \right\} \mapsto \left\{  \frac{\partial g}{\partial \omega_{\rm n}}, \frac{\partial g}{\partial \vep} \right\}$ pullback map for gradient back-propagation through the fixed-$\omega$ eigenproblem.

\section{Optimization of Second Harmonic Conversion Bandwidth}\label{sec:demonstration}

To demonstrate the utility of this design approach we explore the optimization of an etched TFLN waveguide for maximized second harmonic generation (SHG) phase-matching bandwidth. 
We chose this design goal because broadband simultaneous phase matching of optical three-wave mixing in TFLN waveguides is of great interest for a wide range of applications\cite{JankUB, elkus2019generation, fergestad2020ultrabroadband, mishra2022ultra, du2023highly, ledezma2023octave, aashna2023theoretical, sekine2023multi, roy2023visible, williams2024ultra, wu2024visible, huang2024conformal}.
In this example the rib waveguide geometry is parameterized by core width, core thickness and normalized etch fraction. 
The phase mismatch between fundamental and second harmonic quasi-TE$_{00}$ modes is assumed to be compensated by periodic poling of the TFLN \cite{fejerQPM}.
We perform the optimization purely based on the desired group indices and calculate the required poling period from the corresponding phase indices.
This is practical because inverting the TFLN domains has no impact on the linear dielectric polarizability of LiNbO$_3$ and hence the optimized modal dispersion.
To lowest order the bandwidth of the phase-matched SHG process is inversely proportional to the modal group velocity mismatch between first and second harmonic frequencies. 

\begin{align}
    \Delta\omega \propto & \left|n_{g,2\omega} - n_{g,\omega}\right|^{-1} \label{eq:shg_bandwidth}
\end{align}

\subsection{Single Frequency Optimization} 

We first attempted to maximize SHG bandwidth around target optical frequencies by minimizing the magnitude of the modal group index mismatch in Eq.~\ref{eq:shg_bandwidth} at the target frequency.

We defined an objective function equal to the square of the group index difference to avoid discontinuities in the objective function derivative which would be introduced by the absolute value

\begin{align}
    g = & (n_{g,2\omega}-n_{g,\omega})^2.
\end{align}

This objective function is chosen for simplicity and empirically works well. 
We note that $\Delta\omega$ in Eq.~\ref{eq:shg_bandwidth} is limited by higher order dispersion terms when $\left|n_{g,2\omega} - n_{g,\omega}\right|^{-1}$ approaches zero.
Direct optimization over a large bandwidth may be achieved by choosing a different objective function, for example the sum of square group index differences at multiple wavelengths or a combination of group index mismatch and group velocity dispersion mismatch. 

\begin{figure}[htbp]
    \centering{
            \phantomsubcaption\label{fig:optimization:a}
            \phantomsubcaption\label{fig:optimization:b} 
            \phantomsubcaption\label{fig:optimization:c}
            \phantomsubcaption\label{fig:optimization:d}
            \phantomsubcaption\label{fig:optimization:e}
            \phantomsubcaption\label{fig:optimization:f}
            \phantomsubcaption\label{fig:optimization:g}
            \phantomsubcaption\label{fig:optimization:h}
        }
    \includegraphics[width=\textwidth]{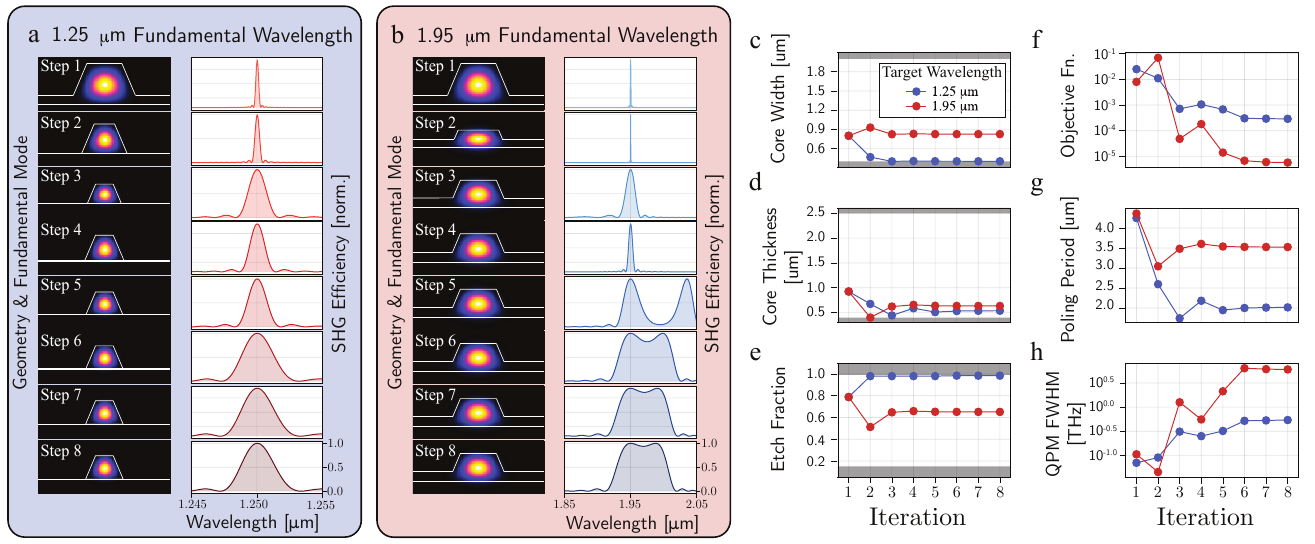}
    \caption{ Evolution of the cross section, fundamental wavelength mode, and effective conversion envelope for 8 iterations at a fundamental wavelength of 
    \subref{fig:optimization:a} \SI{1.25}{\um} and
    \subref{fig:optimization:b} \SI{1.95}{\um}.
    The waveguide was parameterized by its
    \subref{fig:optimization:c} top width (having a $\sim$ \SI{67}{\deg} sidewall angle), 
    \subref{fig:optimization:d} thickness, and 
    \subref{fig:optimization:e} partial etch fraction. 
    0 is un-etched and 1 is fully-etched.
    \subref{fig:optimization:f} The associated objective function values $ (n_{g,2\omega} -n_{g,\omega})^2$.
    \subref{fig:optimization:g} Required poling period to compensate the mismatch in effective index and
    \subref{fig:optimization:h} the resulting FWHM over which the SHG process is quasi phase matched for a \SI{1}{cm} length.
    }
    \label{fig:optimization}
\end{figure}

The frequency dependence and anisotropy of lithium niobate's dielectric susceptibility are crucial factors in this design problem, highlighting the novel capability of our implementation.
Figure~\ref{fig:optimization}\subref{fig:optimization:a} and \ref{fig:optimization}\subref{fig:optimization:b} show the evolution of the waveguide cross section optimization for fundamental wavelengths of \SI{1.25}{\um} and \SI{1.95}{\um}.
The corresponding evolution of parameter and objective function values is shown in Fig.~\ref{fig:optimization}\subref{fig:optimization:c}-\subref{fig:optimization:f} and derived values in Fig.~\ref{fig:optimization}\subref{fig:optimization:g}-\subref{fig:optimization:h}.
Beginning with randomly chosen values for the waveguide geometry, in only 8 iterations the optimizer increases the SHG conversion bandwidth by a factor of 3 (20) around \SI{1.25}{\um} (\SI{1.95}{\um}) compared to bulk dispersion. 
We estimate that comparable optimizations based on exhaustive search require $\sim$100$\times$ as many eigenmode calculations for similar results\cite{JankUB,fergestad2020ultrabroadband,zhang2022ultrabroadband}, even in this low-dimensional parameter space.
We expect this performance gap to grow exponentially with number of design parameters, as discussed in Sec.~\ref{sec:performance}.

We performed similar optimizations of SHG bandwidth and modal GVD for many target vacuum wavelengths spanning \SI{1.25}{\um}--\SI{2.5}{\um} to establish the robustness and generality of our approach for optimization of waveguide dispersion.
These results are presented in Supplement~1.
We note that our SHG bandwidth optimizations centered at \SI{1.95}{\um} (Fig.~\ref{fig:optimization}(b)) and \SI{2.03}{\um} (Fig.~S1(g)) reach \SI{3}{\dB} bandwidths of \SI{78}{\nm} and \SI{70}{\nm} respectively for a \SI{1}{\cm} interaction length, which is comparable to the \SI{110}{\nm} bandwidth recently experimentally demonstrated in a \SI{6}{\mm} exhaustive-search-optimized TFLN waveguide \cite{JankUB} ($\sim$\SI{67}{\nm} for \SI{1}{\cm} length).

\subsection{Broadband Optimization}

\begin{figure}[htbp]
    \centering{
            \phantomsubcaption\label{fig:broadband_designs:a}
            \phantomsubcaption\label{fig:broadband_designs:b} 
            \phantomsubcaption\label{fig:broadband_designs:c}
        }
    \includegraphics[width=0.8\textwidth]{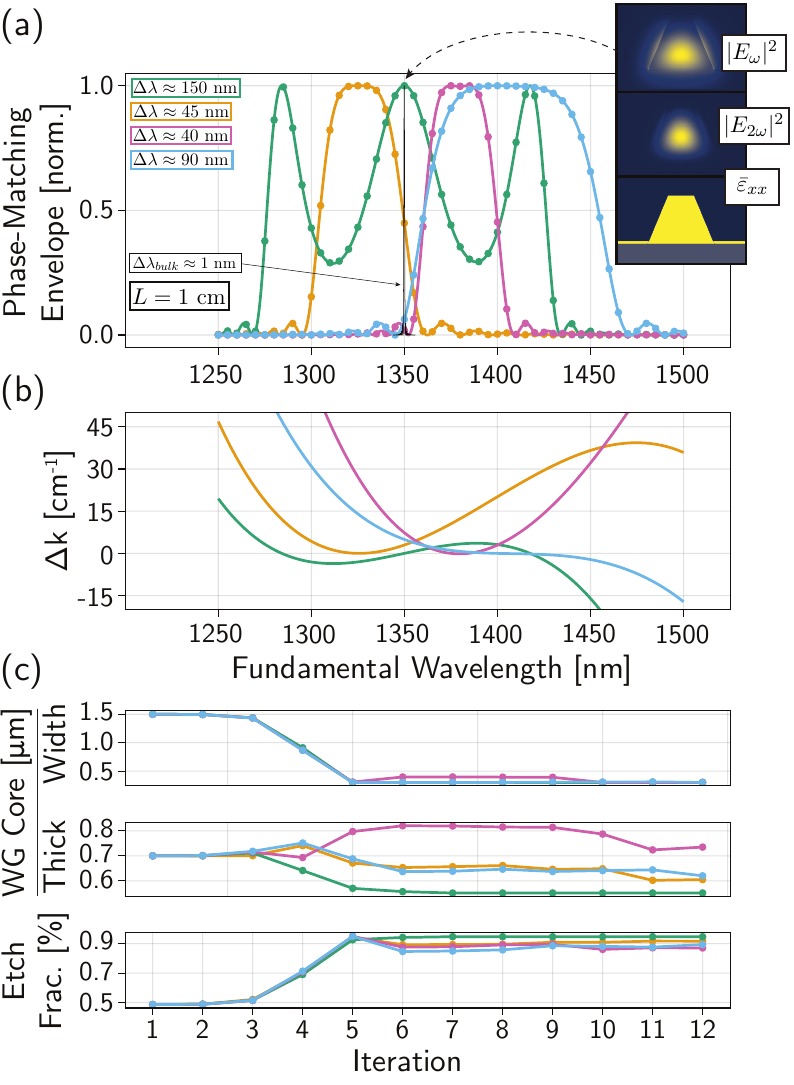}
    \caption{
     \subref{fig:broadband_designs:a} Normalized waveguide SHG transfer functions resulting from multi-frequency waveguide geometry optimization for maximal quasi-phase matching bandwidth around \SI{1325}{\nm} (orange), \SI{1350}{\nm} (green), \SI{1375}{\nm} (pink) and \SI{1400}{\nm} (blue) fundamental wavelengths. In each case the modeled device length is \SI{1}{\cm}. The minimax approach was used to minimize the fundamental/second harmonic group velocity mismatch at several optical frequencies as described in the text. The \SI{1350}{\nm} SHG QPM spectrum in periodically poled bulk LiNbO$_3$ is plotted in black to highlight the $\approx$100$\times$ bandwidth improvement enabled by waveguide dispersion optimization. The TE$_{\rm 00}$ modal intensity distributions for fundamental and second harmonic frequencies are shown in the inset along with a refractive index map showing the optimized waveguide cross-section geometry.
     \subref{fig:broadband_designs:b} SHG phase-mismatch $\Delta k$ spectra corresponding to the optimized SHG transfer function spectra $\propto \textrm{sinc}^2(\Delta k L)$ plotted in \subref{fig:broadband_designs:a}.
     \subref{fig:broadband_designs:c} Parameter traces showing the evolution of the optimized three waveguide design parameters over 12 optimization steps.
    }
    \label{fig:broadband_designs}
\end{figure}

Single-frequency SHG bandwidth optimizations showed limited perforamce at design wavelengths shorter than \SI{1.6}{\um} (see Fig.~S1 in Supplement~1).
In this wavelength range higher order modal dispersion terms tend to be large in magnitude and limit SHG bandwidths to smaller values even when the fundamental and second-harmonic modes have matched group velocities.
To achieve larger SHG QPM bandwidths at shorter wavelengths we re-cast the design problem as a minimax optimization\cite{svanberg2002,boyd2004convex}.
The minimax SHG bandwidth optimization attempts at each step to minimize the largest magnitude group velocity mismatch computed at a discrete set of optical frequencies $\omega_{\rm i}$. 
This approach has been demonstrated to improve optimization results for other types of multi-frequency photonic design problems \cite{Hammond2022High,chen2024validation}.
An epigraph formulation is used to achieve this with a differentiable cost function.
In this form the a dummy variable $t$ is introduced and a set of differentiable inequality constraints $c_{\rm i}(x,t)$ are enforced as the cost function $g(x,t)$ is optimized w.r.t the design parameters $x$.

\begin{equation}
	\begin{gathered}
		g(x,t) = t \\ 
		s.t. \:\:\: c_{\rm i}(x,t) = \left| n_{g,2\omega_{\rm i}}-n_{g,\omega_{\rm i}} \right| - t \leq 0  \label{eq:MiniMaxCostFn}
	\end{gathered}
\end{equation}

We ran separate minimax SHG bandwidth optimizations using sets of three regularly spaced fundamental optical frequencies with center fundamental wavelengths of \SI{1325}{\nm}, \SI{1350}{\nm} , \SI{1375}{\nm} and \SI{1400}{\nm}.
The results of these minimax dispersion optimizations are shown in Fig.~\ref{fig:broadband_designs}.
In each case a broadband phase matching condition is found in the allowed parameter range.
The optimized designs have SHG QPM full-width-half-max bandwidths ranging approximately 40--\SI{150}{\nm} (6--\SI{25}{\THz}) for \SI{1}{\cm} interaction lengths, 30--100$\times$ larger than the corresponding values for bulk PPLN.

Our optimized designs have SHG QPM bandwidths similar to or exceeding previously published results~\cite{JankUB,fergestad2020ultrabroadband,zhang2022ultrabroadband} (6--\SI{25}{\THz} for \SI{1}{\cm} interaction length) at shorter wavelengths where material dispersion makes the design problem more challenging.

The benefit of the multi-frequency approach for broadband dispersion engineering is highlighted in Fig.~\ref{fig:optimization}\subref{fig:optimization:c}.
In single-frequency optimizations shown in Fig.~\ref{fig:optimization}(b) and Fig.~S1 the GVM (the slope of $\Delta k$) at the target wavelength was routinely driven to negligible levels near the precision limit of the simulation ($|n_{g,2\omega} -n_{g,\omega}| \le 10^{-5}$), but the curvature of $\Delta k$ and higher-order dispersion mismatch terms were uncontrolled.
In contrast the minimax-optimized designs reach more moderate GVM magnitudes at the target wavelength ($|n_{g,2\omega} -n_{g,\omega}| \approx 10^{-3}$) but the curvature and higher order dispersion of $\Delta k$ are simultaneously minimized around the target wavelength, further enhancing the simultaneous phase matching bandwidth.
To our knowledge these designs represent the shortest wavelength broadband SHG phase matching conditions predicted in a TFLN platform and the first for the 1.3--\SI{1.4}{\um} wavelength range, where the commercial availability of tunable and swept laser sources~\cite{santecTSLdatasheet,jayaraman2012high} as well as fiber amplification~\cite{thipparapu2016bi,wang2021ultra} make efficient broadband SHG highly desirable.

\section{Performance and Accuracy}\label{sec:performance}

\begin{figure}[htbp]
    \centering{
    \phantomsubcaption\label{fig:scaling:a}
    \phantomsubcaption\label{fig:scaling:b} 
    \phantomsubcaption\label{fig:scaling:c}
    }
    \includegraphics[width=\textwidth]{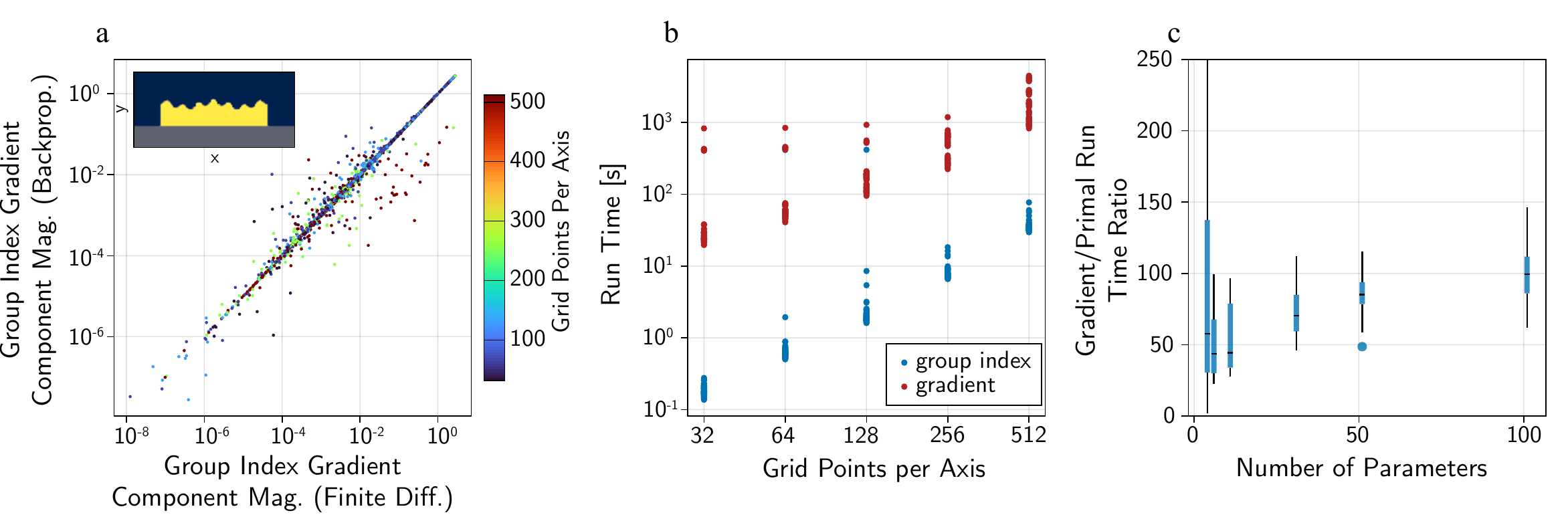}
    \caption{
    \subref{fig:scaling:a} Values of gradient magnitudes as calculated by finite difference and via automatic differentiation. 
    \subref{fig:scaling:b} In the present implementation the gradient calculation dominates total simulation time.
    Scaling for the underlying calculation is driven by DFT computation time for our plane wave expansion-based solver. 
    Run time is given as ``wall time'' on a personal workstation with 32 GB RAM and a 4-core 3.2 GHz Intel processor. 
    \subref{fig:scaling:c} Ratio of time required to compute the gradients vs. the underlying variable (group index or GVD). 
    }
    \label{fig:scaling}
\end{figure}

We performed a series of numerical experiments to study the performance and accuracy of our differentiable mode solver.
We defined a parametric waveguide ridge geometry with a variable number of parameters $N_{\rm p}$ determining a smooth surface morphology, as shown in the inset of Fig.~\ref{fig:scaling:a}.
We then calculated the group index of the highest effective-index waveguide mode and its gradient w.r.t parameters with varied spatial grid resolutions (square grids 32$\times$32\textendash512$\times$512 with fixed spatial extent) and number of geometry parameters $3\le N_{\rm p} \le 100$.
We used 10 randomly generated sets parameter values for each combination of each grid size and number of parameters and separately recorded the calculation times of each primal and gradient calculation.
To assess the gradient accuracy we compared gradients calculated using the adjoint method to nominally equivalent finite difference gradient approximations calculated using a 5-point central difference scheme \cite{FiniteDifferences}.
This comparison is shown in Fig.~\ref{fig:scaling:a}, where each point represents one component of a computed gradient vector and points lying along $y=x$ indicate agreement between adjoint-based and finite-difference gradients.
Some discrepancies are observed but the vast majority of the $\sim$10000 gradient elements compared show good agreement, especially those with the largest magnitude.

The group index and gradient calculation run times (as wall times) are plotted as a function of grid size in Fig.~\ref{fig:scaling:b}.
The run time for the primal group index calculation is dominated by the underlying eigenmode solve.
The observed run-time scaling with respect to grid size appears to be consistent with the $\mathcal{O}(N\mathrm{log}(N))$ scaling (where $N$ is the total number of grid points) seen with the software package \cite{JohnsonMPB} from which our solver is adapted.
In our implementation the gradient calculation is a factor of $\sim$50--100 slower than the primal for all spatial grid sizes and is roughly independent of the number of design parameters $N_{\rm p}$. 
This is not fundamental to our approach and we expect that improvements of the implementation will yield group index/gradient run time ratios of $\sim$1.  

\begin{table}[htbp]
\footnotesize
    \ra{1.3}
    \centering
    \begin{tabular}{ @{}rccccC{2.5cm}@{}}
        \multicolumn{6}{c}{ {\normalsize Table~\ref{tab:comparison}. Comparison of Waveguide Dispersion Engineering Approaches}} \\
        \toprule
        \multicolumn{1}{c}{ \multirow{ 2}{*}{Work} }    &   \multirow{ 2}{*}{\parbox{1.7cm}{\centering Optimization Approach}} &  \multirow{ 2}{1.5cm}{\centering Mode Solves Per Iter. $[N]$}   &   \multicolumn{2}{c}{\centering Materials } & \multirow{2}{2.3cm}{\centering Geometry Definition} \\
            \cmidrule{4-5}
         &  &   & Anisotropy & Dispersion  \\
        \hline
        Talenti, \emph{et al}. \cite{talentiKdotP} & $k\cdot p$ reduced model & $\mathcal{O}(1)^a$  & No & No & 1-parameter PhC \\
        Castell{\'o}, \emph{et al}. \cite{CastelloSiWG} & Hellmann-Feynman &  $\mathcal{O}(1)$	  & No    & No &  Limited Param. \\
        Minkov, \emph{et al}. \cite{MinkovPhC} 	&  Auto-Diff. & $\mathcal{O}(1)$	& No 	& No	& Limited Param. \\
        Vial and Hao \cite{vial2022autodiff}                &  Auto-Diff. &  $\mathcal{O}(1)$   &   Partial$^b$    &  No &  Binary Topology \\
        Hameed, \emph{et al}. \cite{hameedTrustRegion} & Trust Region Algo.  & $\mathcal{O}(N^3)$ & Yes & No & Limited Param.  \\
        \multicolumn{1}{c}{\textbf{This Work}} &  Auto-Diff.  &  $\mathcal{O}(1)$ & Yes & Yes	& Arbitrary Param. \\ 
        \bottomrule
    \end{tabular}	
    \caption{ Abbreviations: Automatic differentiation (``Auto-Diff.''), Trust-Region Algorithm (``Trust Region Algo.''), Limited parameterization (``Limited Param.''), Arbitrary Parameterization (``Arbitrary Param.''), Photonic Crystal (``PhC''), Hellmann-Feynman Theorem (``Hellmann-Feynman''). 
    ``Limited parameterization'' geometries are, for example, the position and size of holes in a PhC or the width and height of a rectangular waveguide. 
    In contrast, ``arbitrary parametization'' allows for any geometrical shape which can be described by a finite number of input parameters. 
    We describe ``limited'' and ``arbitrary'' parametrizations based on what is reported in each work; these may not be fundamental to the approach.
    Additionally, we consider inclusion of ``dispersion'' to be explicit use of material information to determine spectral derivatives without finite differences in frequency. \\
    $^a$Only geometries parameterized in 1 variable are supported, then $\mathcal{O}(1)$ 3D FDTD solves are required; additionally $\sim100$s of (fast) reduced model solves are performed. \\
    $^b$Anisotropic materials supported for uniaxial materials with z-oriented anisotropy.
    }
    \label{tab:comparison}
\end{table}

The ratios between corresponding group index and gradient calculation run times are plotted as a function of the number of design parameters $N_{\rm p}$ in Fig.~\ref{fig:scaling:c}.
We observe that the ratio between gradient and primal calculation times approximately doubles for a 30-fold increase in $N_{\rm p}$. 
This indicates that the gradient calculation run-time does not substantially depend with $N_{\rm p}$, as expected for gradients computed using the adjoint method.

We computed independent eigen- and adjoint problem solutions in parallel for objective functions depending on multiple waveguide mode solutions sharing a common geometry (for example modes at several distinct optical frequencies).
Thus using a parallel computing environment we were able to scale the spectral bandwidth or resolution of multi-frequency objective functions without appreciably increasing simulation time.

Table~\ref{tab:comparison} compares this work with several previous demonstrations of inverse design for dispersion engineering.
While many are based on similar frameworks, none support the dispersive and anisotropic materials, limiting their relevance to key nonlinear optical applications. 
Furthermore, our approach back-propagates gradients w.r.t both the eigenvalues \emph{and} eigenvectors (mode fields), a limitation of \cite{CastelloSiWG}.

\section{Conclusion}\label{sec:conclusion}

In summary we have demonstrated a differentiable electromagnetic eigenmode solver suitable for a wide variety of modeling and design applications and applied it to optimize waveguide geometries for maximal SHG phase matching bandwidth.
We demonstrated optimized designs for broadband SHG in TFLN waveguides in the 1.3--\SI{1.4}{\um} wavelength range, enabling widely tunable visible light sources based on commercially available infrared lasers.
Our results for a practical waveguide dispersion optimization with three parameters demonstrate a $\sim$100-fold reduction in the number of required mode solutions compared to commonly used exhaustive search approaches.
To enable this we have derived the adjoint problem and back-propagation equations for waveguide mode solutions accounting for frequency-dependent and anisotropic dielectric materials and sub-pixel smoothing. 
We believe this is the first mode solver capable of back-propagating gradients with respect to mode fields and effective indices, group indices and GVDs calculated from single eigenmode solutions.
In this work we focused on 2D waveguide cross-section models, however the solver and the adjoint method described in our paper also applies to 3D eigenmode models.

This type of differentiable mode solver can be modularly combined with FDTD, eigenmode expansion (EME) and multi-physics models to enable optimizable end-to-end models of integrated photonic and optoelectronic components and systems.
Our work enables application of inverse design methods to a wide variety of integrated photonic design problems based on waveguide mode models.

\begin{backmatter}

\bmsection{Funding}
This research was supported in part by an appointment to the Intelligence Community Postdoctoral Research Fellowship Program at the Massachusetts Institute of Technology administered by Oak Ridge Institute for Science and Education (ORISE) through an interagency agreement between the U.S. Department of Energy and the Office of the Director of National Intelligence (ODNI), and in part by the DARPA DSO Robust Optical Clock Netowrk (ROCkN) Program.

\bmsection{Acknowledgments}
The authors would like to thank Steven G. Johnson for helpful discussions. 
The authors acknowledge the MIT SuperCloud and Lincoln Laboratory Supercomputing Center for providing HPC resources that have contributed to the research results reported within this paper.

\bmsection{Disclosures}
The authors declare no conflicts of interest.

\bmsection{Data Availability Statement}
The presented data is available upon reasonable request.

\bmsection{Supplemental document}
See Supplement 1 for supporting content. 

\end{backmatter}

\section{References}
\label{sec:refs}
\bibliography{references}

\end{document}


\maketitle

\section{Introduction}

In the following we describe the differentiable electromagnetic mode solver implemented for this work. 
The forward (primal) calculation of waveguide mode fields and wave numbers closely follows \cite{JohnsonMPB}. 
The adjoint method for sensitivity analysis of eigen-mode solutions follows \cite{johnsonAdj}.
The modified adjoint problem for the case of a frequency-dependent dielectric tensor and the equations relating frequency and dielectric tensor element sensitivities to the adjoint field were derived as part of this work.

As discussed in the main text, sub-pixel smoothing of the dielectric tensor near material interfaces ($\varepsilon \rightarrow \tilde{\varepsilon}$) is required for models of waveguide modes to be smoothly differentiable with respect to geometry parameters \cite{KottkeSPS,OskooiSmoothing}. 
The following sensitivity analysis of mode fields and modal dispersion with respect to dielectric tensor element values does not require or assume dielectric smoothing. 
Here we will refer to the dielectric tensor data and its frequency derivatives as $\varepsilon$, $\frac{d\varepsilon}{d\omega}$ and $\frac{d^2\varepsilon}{(d\omega)^2}$.
The calculations presented in the main text used smoothed tensor data for these variables.

We first consider the eigenmode solution and gradient back-propagation for the case of a frequency-independent dielectric function in Sections \ref{SecLinearEigenproblem} and \ref{SecLinearEigenproblemAdj}, respectively.
We generalize this result to the case where dielectric function is frequency-dependent in Sections \ref{sec:sup:SecNonlinearEigenproblem} and \ref{sec:sup:NonlinearEigenproblemAdj}, respectively.

\section{The Helmholtz Eigenproblem in a Plane Wave Basis}\label{SecLinearEigenproblem}

We consider a Helmholtz Equation for the magnetic field $H_{\rm n}$ of the $n$-th electromagnetic waveguide mode propagating in the $+\hat{z}$ direction (along the waveguide axis), parameterized by dielectric tensor values $\vep(x,y,z)$ at each point on a regular spatial grid $ \left\{ x_{\rm i},y_{\rm i},z_{\rm i} \right\} $ and a wave vector $k\hat{z}$ with magnitude $k$

    \begin{equation}
        \mathbf{M}(k,\vep) H_{\rm n} = \omega_{\rm n}^2 H_{\rm n} \label{HelmholtzEq}
    \end{equation}

The Helmholtz operator $\mathbf{M}$ consists of two curl operations sandwiching multiplication by the inverse dielectric tensor $\vep^{-1}$. 

    \begin{equation}
        \mathbf{M}(k,\vep)  = \left[\nabla \times\right]  \left[\vep^{-1} \cdot \right]  \left[\nabla \times\right] \label{HelmholtzOp}
    \end{equation}

In this work we focus on waveguide modes propagating along $\hat{z}$ and confined in the transverse dimensions, but the analysis that follows is straightforward to generalize to any eigenmode solutions of the electromagnetic Helmholtz Eqaution.
For reasons discussed below it is most convenient to solve Eq. \ref{HelmholtzOp} with periodic boundary conditions, despite the fact that we are interested in dielectric waveguide geometries which are not periodic in the dimensions transverse to the propagation axis (here the $x-y$ plane). 
In practice we can approximate dielectric waveguide geometries which are not periodic in the transverse dimensions by simulating sufficiently large spatial extents $Dx$ and $Dy$ that evanescent fields decay to negligible levels at the $x-$ and $y-$ boundaries and by including vacuum gaps to stifle coupling across transverse boundaries.

The Helmholtz operator can be efficiently applied to mode field data by computing the curl operations in a plane wave basis (i.e. in reciprocal space). 
The change of basis between real-space field data with Cartesian vector components to plane wave basis field data with transverse vector components is comprised of two steps. 
First a discrete Fourier transform (DFT) is performed on each Cartesian vector component, resulting in a reciprocal space Cartesian vector field $\Tilde{H}_{\rm j}$ at each reciprocal lattice point $\vec{g}_{\rm j}$

    \begin{eqnarray}
        \Tilde{H} &=& \mathbf{F} \cdot H \label{EqDFTOp1} \\
        \Tilde{H}_{\rm j} &=& \sum_{\rm i} \mqty[ H_{\rm i,x} \\  H_{\rm i,y} \\ H_{\rm i,z} ] e^{ 2 \pi i \vec{g}_{\rm j} \cdot \vec{x}_{\rm i} } \label{EqDFTOp2}
    \end{eqnarray}

where $\mathbf{F}$ in Eq.~\ref{EqDFTOp1} represents the DFT linear operator specified by Eq.~\ref{EqDFTOp2}. 
Next the Cartesian reciprocal space field $\Tilde{H}_{\rm j}$ is projected onto a pair of orthonormal  basis vectors $\left\{\hat{m}_{\rm j},\hat{n}_{\rm j}\right\}$ transverse to each wave vector $\vec{v}_j$ in the plane wave basis

    \begin{equation}
        \vec{v}_j = k\hat{z} + \vec{g}_j \label{eqn:VjDef}
    \end{equation}

    \begin{equation}
        \Tilde{h}_j = \mqty[ \hat{m}_{j,\rm x} & \hat{m}_{j,\rm y} & \hat{m}_{j,\rm z}  \\  \hat{n}_{j,\rm x} & \hat{n}_{j,\rm y} & \hat{n}_{j,\rm z}  ] \mqty[ \Tilde{H}_{j,\rm x}  \\  \Tilde{H}_{j,\rm y} \\ \Tilde{H}_{j,\rm z} ] \label{EqCartToTransProjOp}
    \end{equation}

The transverse basis vectors $\left\{\hat{m}_{\rm j},\hat{n}_{\rm j}\right\}$ are chosen systematically

    \begin{eqnarray}
        \hat{n}_j &=& \frac{\vec{a} \times \vec{v}_j}{\left| \vec{a} \times \vec{v}_j \right|} \label{EqKtoN} \\
        \hat{m}_j &=& \frac{\hat{n}_j \times \vec{v}_j}{\left| \hat{n}_j \times \vec{v}_j \right|} \label{EqKtoM}
    \end{eqnarray}

for some arbitrary Cartesian seed vector $\vec{a}$. The $k \mapsto \left\{\hat{m}_j,\hat{n}_j\right\}$ mapping in Eqs.~\ref{EqKtoN}~\&~\ref{EqKtoM} is smooth and differentiable w.r.t $k$ as long as $\vec{a} \nparallel \vec{v}_j$, the so-called ``gimbal lock'' case which we ignore for now. The curl acts locally in reciprocal space, swapping and scaling each pair of transverse vector components 
    
    \begin{eqnarray}
        \nabla \times \Tilde{h}_j &=& \vec{v}_j \times \Tilde{h}_j \\
            &=&  \begin{bmatrix} 0 & \left| \vec{v}_j \right| \\ -\left| \vec{v}_j \right| & 0 \end{bmatrix} \mqty[ \Tilde{h}_{j,\hat{m}} \\ \Tilde{h}_{j,\hat{n}} ] \label{EqPWCurlOp}
    \end{eqnarray}

Eq.~\ref{HelmholtzEq} can be expressed in the plane wave basis as 

    \begin{equation}
        \left[ \mathbf{C}^\dagger \vep^{-1}  \mathbf{C} \right] \Tilde{h}_{\rm n} = \omega^2_{\rm n} \Tilde{h}_{\rm n} \label{PWHelmholtzEq}
    \end{equation}

where $\mathbf{C}$ represents the fused DFT, Cartesian-to-transverse projection, and curl operators defined in Eqs.~\ref{EqDFTOp2},~\ref{EqCartToTransProjOp}~\&~\ref{EqPWCurlOp}, respectively.

    \begin{eqnarray}
        \mathbf{C} &=&  \mathbf{F}  \mqty[ \hat{m}_j & \hat{n}_j ]   \begin{bmatrix} 0 & \left| \vec{v}_j \right| \\ -\left| \vec{v}_j \right| & 0 \end{bmatrix} \label{EqCkDef1} \\
        &=& \mathbf{F}  \left| \vec{v}_j \right| \mqty[ - \hat{n}_j & \hat{m}_j ]  \label{EqCkDef}
    \end{eqnarray}

Approximations of one or more electromagnetic modes of a dielectric structure described by $\vep$ can be found using an iterative algorithm for partial eigen-decomposition of Eq.~\ref{PWHelmholtzEq}, such as a conjugate gradient method. For the purpose of dispersion optimization it is useful to view this partial eigen-decomposition as a mapping $ \left\{ k, \vep \right\} \mapsto \left\{ \omega_{\rm n}, \Tilde{h}_{\rm n} \right\} $. Away from modal degeneracies where $\omega_{\rm n}(k) = \omega_{\rm m}(k) $ for some mode indices $m \neq n$, this mapping is smooth and differentiable.

\section{ Gradient Back-Propagation through Solutions of the Linear\\*Helmholtz~Equation}\label{SecLinearEigenproblemAdj}

We now consider the case where an electromagnetic mode of interest with frequency $\omega_{\rm n}$ and mode field $\Tilde{h}_{\rm n}$ has been found as described in Sec.~\ref{SecLinearEigenproblem} and a downstream objective function $g(...,\omega_{\rm n}, \Tilde{h}_{\rm n},...)$ depending on $\omega_{\rm n}$ and $\Tilde{h}_{\rm n}$ has been computed, along with the gradient components of the objective function w.r.t $\omega_{\rm n}$ and $\Tilde{h}_{\rm n}$, $\frac{\partial g}{\partial \omega_{\rm n}}$ and $\frac{\partial g}{\partial \Tilde{h}_{\rm n}}$. 
This last step, commonly referred to as ``back-propagation'' in the machine learning literature, can be accomplished with manually written gradient functions or by automatic differentiation (AD). 
$\frac{\partial g}{\partial \omega_{\rm n}}$ and $\frac{\partial g}{\Tilde{h}_{\rm n}}$ can be efficiently mapped to gradients w.r.t the parameters determining Eq.~\ref{PWHelmholtzEq}, $\frac{\partial g}{\partial \vep}$ and $\frac{\partial g}{\partial \vec{k}}$, using the adjoint-field method \cite{johnsonAdj}. 
The plane-wave-basis adjoint field $\Tilde{h}_{\lambda \rm n}$ corresponding to gradients w.r.t the $\mathrm{n^{th}}$ mode solution $\frac{\partial g}{\partial \Tilde{h}_{\rm n}}$ and $\frac{\partial g}{\partial \omega^2_{\rm n}}$ is calculated according to 

    \begin{equation}
        \Tilde{h}_{\lambda \rm n}    =         \Tilde{h}_{\lambda \rm n 0} + \frac{\partial g}{\partial \omega^2_{\rm n}} \Tilde{h}_{\rm n} 
    \end{equation}

where $\Tilde{h}_{\lambda \rm n 0}$ is the solution to

    \begin{equation}
        \left[\mathbf{M} -  \omega_{\rm n}^2\mathbf{I}\right] \Tilde{h}_{\lambda \rm n 0} = \left[ \mathbf{I} - \Tilde{h}_{\rm n} \Tilde{h}_{\rm n}^\dagger \right] \left( \frac{\partial g}{\partial \Tilde{h}_{\rm n}} \right)^\dagger \label{AdjFieldEqn}
    \end{equation}

Eq.~\ref{AdjFieldEqn} can be solved approximately using a slightly modified version of the same ``matrix-free'' implementation of $\mathbf{M}$ used to solve for $\Tilde{h}_{\rm n}$ and $\omega_{\rm n}^2$ and an iterative method (eg. GMRES).
We note that the bracketed term on the right-hand-side of Eq.~\ref{AdjFieldEqn} is a projection onto the space of plane wave basis mode fields orthogonal to $\Tilde{h}_{\rm n}$ (also the null space of $\left[\mathbf{M} -  \omega_{\rm n}^2\mathbf{I}\right]$). 

The outer product of $\Tilde{h}_{\lambda \rm n}$ with $\Tilde{h}_{\rm n}$ gives the gradient of the objective function w.r.t matrix elements of the Helmholtz operator $\mathbf{M}$ 

    \begin{equation}
         \frac{\partial g}{\partial \mathbf{M}} = \Tilde{h}_{\lambda \rm n} \Tilde{h}_{\rm n}^\dagger \label{EqMbar}
    \end{equation}

Rather than explicitly computing $\frac{\partial g}{\partial \mathbf{M}}$, we can use Eq.~\ref{EqMbar} to derive formulas for $\frac{\partial g}{\partial k}$ and $\frac{\partial g}{\partial \vep}$ in terms of $\Tilde{h}_{\lambda \rm n}$. The gradients for the operator components of $\mathbf{M}$ as written in Eq.~\ref{PWHelmholtzEq} can be computed from standard matrix-derivative results\cite{giles2008collected}.

First we compute $\frac{\partial g}{\partial \vep}$ 
    \begin{eqnarray}
        \frac{\partial g}{\partial \vep} &=& \vep^{- \dagger} \mathbf{C} \frac{\partial g}{\partial \mathbf{M}}  \mathbf{C}^\dagger \vep^{- \dagger} \\
        &=& \vep^{- \dagger} \mathbf{C} \Tilde{h}_{\lambda \rm n} \Tilde{h}_{\rm n}^\dagger  \mathbf{C}^\dagger \vep^{- \dagger} \\
        &=& E_{\rm \lambda n} E_{\rm n}^\dag \label{EpsilonGrad}
    \end{eqnarray}

where $E_{\rm n}$ is the real space modal electric field corresponding to $\Tilde{h}_{\rm n}$

\begin{eqnarray}
    E_{\rm n}   &=& \vep^{-1} \mathbf{C} \Tilde{h}_{\rm n} \label{EqEFieldDef} \\
                &=& \vep^{-1} \left[ \nabla \times H_{\rm n} \right]
\end{eqnarray}

and $E_{\rm \lambda n}$ is the analogous ``adjoint electric field'' corresponding to $\Tilde{h}_{\lambda \rm n}$

\begin{equation}
    E_{\rm \lambda n}   = \vep^{-\dagger} \mathbf{C} \Tilde{h}_{\lambda \rm n} \label{EqAdjFieldDef}
\end{equation}

Note the dielectric tensor operator in Eqn.~\ref{EqAdjFieldDef} is the adjoint of the corresponding term in Eqn.~\ref{EqEFieldDef} for the modal electric field. This is irrelevant for models with Hermitian $\vep$ which neglect dielectric loss or gain.

Next we compute $\frac{\partial g}{\partial \mathbf{C}}$

\begin{eqnarray}
    \frac{\partial g}{\partial \mathbf{C}} &=& \vep^{-1} \mathbf{C}\left(\frac{\partial g}{\partial \mathbf{M}}\right)^\dagger + \vep^{-\dagger}\mathbf{C}\frac{\partial g}{\partial \mathbf{M}} \\
    &=& \vep^{-1} \mathbf{C} \Tilde{h}_{\rm n}\Tilde{h}_{\lambda \rm n}^\dagger + \vep^{-\dagger}\mathbf{C}\Tilde{h}_{\lambda \rm n} \Tilde{h}_{\rm n}^\dagger \label{CkGrad} \\
    &=& E_{\rm n} \Tilde{h}_{\lambda \rm n}^\dag + E_{\rm \lambda n} \Tilde{h}_{\rm n}^\dag \label{CkGradHerm}
\end{eqnarray}

We back-propagate $\frac{\partial g}{\partial \mathbf{C}}$ gradient components to $\frac{\partial g}{\partial |\vec{v}_j|}$ and $\frac{\partial g}{\partial [ \hat{m}_j, \hat{n}_j ]}$ using the definition of $\mathbf{C}$ (Eqn.~\ref{EqCkDef})

\begin{equation}
    \frac{\partial g}{\partial [ \hat{m}_j, \hat{n}_j ]} = \left[ \Tilde{E}_{\rm n} \Tilde{h}_{\lambda \rm n}^\dag + \Tilde{E}_{\rm \lambda n} \Tilde{h}_{\rm n}^\dag  \right] \begin{bmatrix} 0 & -\left| \vec{v}_j \right| \\ \left| \vec{v}_j \right| & 0 \end{bmatrix} \label{eq:sup:dLdmn}
\end{equation}

\begin{equation}
    \frac{\partial g}{\partial |\vec{v}_j|} = \Tr \left( \left[ \Tilde{E}_{\rm n} \Tilde{h}_{\lambda \rm n}^\dag + \Tilde{E}_{\rm \lambda n} \Tilde{h}_{\rm n}^\dag  \right] \mqty[ -\hat{n}_j & \hat{m}_j ]^T \right) \label{eq:sup:dLdv}
\end{equation}

Finally $\frac{\partial g}{\partial k}$ is computed by back-propagating $\frac{\partial g}{\partial |\vec{v}_j|}$ and $\frac{\partial g}{\partial [ \hat{m}_j, \hat{n}_j ]}$ through Eqs.~\ref{eqn:VjDef}, \ref{EqKtoN} \& \ref{EqKtoM} and adding $\frac{\partial g}{\partial \omega_{\rm n}^2} \frac{\partial \omega^2_{\rm n}}{\partial k}$

\begin{equation}
    \frac{\partial g}{\partial k} = \frac{\partial g}{\partial \omega_{\rm n}^2} \frac{\partial \omega^2_{\rm n}}{\partial k} + \sum_j \Tr\left( \mqty[ -(v_{j\rm z}\hat{n}_j + n_{j\rm z} \vec{v}_j)^T  \\  (v_{j\rm z}\hat{m}_j + m_{j\rm z} \vec{v}_j)^T ] \left[ \Tilde{E}_{\rm n} \Tilde{h}_{\lambda \rm n}^\dag + \Tilde{E}_{\rm \lambda n} \Tilde{h}_{\rm n}^\dag \right]\right) \label{KGrad}
\end{equation}

with $\frac{\partial \omega^2_{\rm n}}{\partial k} = 2 \omega_{\rm n} \frac{\partial \omega_{\rm n}}{\partial k} = 2 \frac{\omega_{\rm n}}{n_{\rm g, n}}$ computed according to the Hellmann-Feynman theorem\cite{JohnsonMPB}

\begin{eqnarray}
   \frac{\partial \omega^2_{\rm n}}{\partial k} &=& \frac{\sum \Tilde{h}_{\rm n}^\dagger \frac{\partial \mathbf{M}}{\partial k} \Tilde{h}_{\rm n}}{\sum \Tilde{h}_{\rm n}^\dagger \Tilde{h}_{\rm n}} \\
   &=& \frac{ 2 \sum \Tilde{h}_{\rm n}^\dagger \left[ \mathbf{C}^\dagger \vep^{-1}  \frac{d \mathbf{C}}{dk} \right] \Tilde{h}_{\rm n} }{\sum \Tilde{h}_{\rm n}^\dagger \Tilde{h}_{\rm n}} \label{eqn:HellmannFeynman}     
\end{eqnarray}

where $\frac{d \mathbf{C}}{dk}$ is similar to $\mathbf{C}$ (Eq.~\ref{EqCkDef}) but with the curl operator $\left[\vec{k}+g_{\rm j}\right]\times$ replaced by $\hat{z}\times$ 

\begin{equation}
    \frac{d \mathbf{C}}{dk} = \mathbf{F} \begin{bmatrix} 0 & 1 & 0 \\ -1 & 0 & 0 \\ 0 & 0 & 0 \end{bmatrix} \mqty[ \hat{m}_{\rm j} & \hat{n}_{\rm j} ]    \label{EqBzDef}
\end{equation}

Equations~\ref{EpsilonGrad} \& \ref{KGrad} provide the pullback map $\left\{ \frac{\partial g}{\partial \omega_{\rm n}}, \frac{\partial g}{\partial \Tilde{h}_{\rm n}} \right\} \mapsto \left\{ \frac{\partial g}{\partial k}, \frac{\partial g}{\partial \vep} \right\}$ for back-propagating gradient components through the the mode solution map $ \left\{ k, \vep \right\} \mapsto \left\{ \omega_{\rm n}, \Tilde{h}_{\rm n} \right\} $. 
Defining a function computing this pullback map as a custom rule for an AD framework enables reverse-mode differentiation of arbitrary objective functions that depend on waveguide mode fields and effective indices.
As usual for adjoint method sensitivity analysis, the cost (per mode) of back-propagation through the mode solver is similar to the cost of finding that mode on the forward pass and does not scale with the number of parameters.

\section{Nonlinear (fixed-$\omega$) Eigenproblem}\label{sec:sup:SecNonlinearEigenproblem}

In this work we are concerned with the value and derivatives of the modal eigenvalue dispersion ($\frac{dk}{d\omega}$, $\frac{d^2k}{d\omega^2}$,...). 
In practice we need to account for the frequency dependence of the dielectric tensor $\vep(\omega)$ to compute these quantities accurately over frequency ranges comparable to the optical frequency itself ($\frac{\Delta\omega}{\omega}\approx1$). 
When the dielectric tensor $\vep$ is frequency-dependent, the eigenproblem in Eq.~\ref{HelmholtzEq} becomes nonlinear.

    \begin{equation}
        \mathbf{M}\left(k,\vep(\omega_{\rm n}) \right) H_{\rm n} = \omega_{\rm n}^2 H_{\rm n} \label{NonlinearHelmholtzEqn}
    \end{equation}

Eq.~\ref{NonlinearHelmholtzEqn} can be solved iteratively for the unknown wave vector magnitude $k$ with fixed eigenvalue $\omega^2$ and dielectric tensor $\vep(\omega)$ using the Newton method. 
The $k$ approximation is updated after each Newton step according to 

    \begin{equation}
        k_{\rm i+1} = k_{\rm i} - \left.\frac{\partial \omega^2}{\partial k}\right|_\vep^{-1} (\omega^2 - \omega_{\rm i}^2)
    \end{equation}

where the eigenvalue gradient w.r.t $k$ for fixed $\vep$, $\left.\frac{\partial \omega^2}{\partial k}\right|_\vep^{-1}$, is calculated using Eqn.~\ref{eqn:HellmannFeynman}.
The solution of this nonlinear eigenproblem accounting for dielectric dispersion maps $\{\omega,\vep\} \mapsto \{k_{\rm n}, \tilde{h}_{\rm n}\}$.

\section{Gradients for Nonlinear (fixed-$\omega$) Eigenproblem}\label{sec:sup:NonlinearEigenproblemAdj}

Equation~\ref{NonlinearHelmholtzEqn} can be understood as the combination of Eq.~\ref{HelmholtzEq} with an additional constraint enforcing equality between the n-th eigenvalue $\omega_{\rm n}^2$ and the square of a target input frequency $\omega$.

    \begin{equation}
        f(k,\omega,\vep(\omega)) = \mathrm{eigvals}\left(\mathbf{M}\left(k,\vep(\omega)\right)\right)_{\rm n} - \omega^2 = 0 \label{eq:sup:EqEigenvalueConstraint}
    \end{equation}

Differentiating Eq.~\ref{eq:sup:EqEigenvalueConstraint} gives the following relationships between partial derivatives

    \begin{eqnarray}
        \frac{\partial k}{\partial \vep} &=& -\left( \frac{\partial f}{\partial k} \right)^{-\dagger} \frac{\partial f}{\partial \vep} \\
        &=& \frac{\partial k}{\partial \omega^2} \frac{\partial \omega^2}{\partial \vep} \\ 
        &=& \frac{\partial k}{\partial \omega^2} E_{\rm n} E_{\rm n}^\dag \label{EqDkDepsNL}
    \end{eqnarray}

    \begin{eqnarray}
        \frac{\partial g}{\partial \omega^2} &=&  \frac{\partial g}{\partial k} \frac{\partial k}{\partial \omega^2} + \frac{\partial g}{\partial \tilde{h}_{\rm n}} \frac{\partial \tilde{h}_{\rm n}}{\partial \omega^2} \\
        &=& \left[ \frac{\partial g}{\partial k}  + \frac{\partial g}{\partial \tilde{h}_{\rm n}} \frac{\partial \tilde{h}_{\rm n}}{\partial k}\right] \frac{\partial k}{\partial \omega^2} \label{EqDgDomsqNL}
    \end{eqnarray}

    \begin{eqnarray}
        \frac{\partial g}{\partial \vep} &=& \frac{\partial g}{\partial k} \frac{\partial k}{\partial \vep} + \frac{\partial g}{\partial \tilde{h}_{\rm n}} \frac{\partial \tilde{h}_{\rm n}}{\partial \vep} \\
        &=& \frac{\partial g}{\partial k} \left( \frac{\partial k}{\partial \omega^2} \frac{\partial \omega^2}{\partial \vep} \right) + \frac{\partial g}{\partial \tilde{h}_{\rm n}} \left( \frac{\partial \tilde{h}_{\rm n}}{\partial k} \frac{\partial k}{\partial \omega^2}   \frac{\partial \omega^2}{\partial \vep}  \right) \\
        &=&  \left[ \frac{\partial g}{\partial k}  + \frac{\partial g}{\partial \tilde{h}_{\rm n}} \frac{\partial \tilde{h}_{\rm n}}{\partial k}\right] \frac{\partial k}{\partial \omega^2} \frac{\partial \omega^2}{\partial \vep} \\
        &=&  \frac{\partial g}{\partial \omega^2} \frac{\partial \omega^2}{\partial \vep}  \label{EqDgDepsNL}
    \end{eqnarray}

where $\frac{\partial g}{\partial \tilde{h}_{\rm n}} \frac{\partial \tilde{h}_{\rm n}}{\partial k}$ in Eq.~\ref{EqDgDomsqNL} is computed using the same adjoint method described for the fixed-$k$ case and is equivalent to Eq.~\ref{KGrad} with the first term ($\frac{\partial g}{\partial \omega^2_{\rm n}} \frac{\partial \omega^2_{\rm n}}{\partial k}$) set to zero.
Equations~\ref{EqDgDomsqNL}~\&~\ref{EqDgDepsNL} provide the pullback map for the fixed-$\omega$ eigenproblem $ \left\{ \frac{\partial g}{\partial k}, \frac{\partial g}{\partial \Tilde{h}_{\rm n}} \right\} \mapsto \left\{  \frac{\partial g}{\partial \omega_{\rm n}}, \frac{\partial g}{\partial \vep} \right\}$ in terms of the pullback map for the fixed-$k$ eigenproblem (Equations~\ref{EpsilonGrad}~\&~\ref{KGrad}).

\section{Modal Group Index}\label{sec:sup:group_index}

We calculate the modal group index as the ratio of the integrated modal Poynting flux along the direction of propagation (here $\hat{z}$) $P=\int_A \vec{S}\cdot\hat{z}\;\text{d}A$ and the modal electromagnetic energy density per unit length $W=\int_A U \;\text{d}A$ \cite{jacksonEM}
    \begin{equation}
        n_{\rm g, n} = \frac{c}{v_{\rm g, n}} = \frac{\partial|k_{\rm n}|}{\partial \omega} = \frac{W}{P} = \frac{\int_A U \;\text{d}A}{\int_A \vec{S}\cdot\hat{z} \;\text{d}A} \label{eq:sup:group_index}
    \end{equation}
Where $\vec{S}$ is the Poynting flux density 
    \begin{equation}
        \vec{S} = \: \vec{E} \times \vec{H}
    \end{equation}
and $U$ is the electromagnetic energy density including frequency-dependent dielectric and magnetic susceptibilities 
    \begin{align}
        U = & \: \frac{1}{2}\vec{E}\cdot \frac{\partial (\omega\bar{\vep})}{\partial\omega} :\vec{E} + \frac{1}{2}\vec{H} \frac{\partial (\omega\bar{\mu})}{\partial\omega} : \vec{H}. \\
    \end{align}
$\vec{S}$ and $U$ can be calculated at each real space grid point from the modal electric and magnetic fields. Comparing with Eq.~\ref{eqn:HellmannFeynman}, we identify the sums in the discretized plane wave basis corresponding to these integrals 
    \begin{equation}
        P = \omega_{\rm n} \sum \Tilde{h}_{\rm n}^\dagger \frac{\partial \mathbf{M}}{\partial k} \Tilde{h}_{\rm n}
    \end{equation}

    \begin{align}
        W   = & \sum \Tilde{h}_{\rm n}^\dagger \mathbf{C}^\dagger \vep^{-1} \left[ \vep + \frac{\omega_{\rm n}}{2} \frac{d\vep}{d\omega} \right] \vep^{-1} \mathbf{C} \Tilde{h}_{\rm n} \label{eq:sup:pw_linear_energy_density1} \\
            = & \omega_{\rm n}^2 + \frac{\omega_{\rm n}}{2}\sum \Tilde{h}_{\rm n}^\dagger \left[ \mathbf{C}^\dagger \vep^{-1} \frac{d\vep}{d\omega} \vep^{-1} \mathbf{C} \right] \Tilde{h}_{\rm n} \label{eq:sup:pw_linear_energy_density}
    \end{align}
    
Where in Eqns.~\ref{eq:sup:pw_linear_energy_density1}~\&~\ref{eq:sup:pw_linear_energy_density} we assume the time-averaged electric and magnetic energy integrated over the mode are equal ($\int\vec{E}\cdot\vec{D}=\int\vec{H}\cdot\vec{B}$) and that the frequency dependence of the magnetic susceptibility is negligible ($\frac{d\mu}{d\omega}=0$). 
Thus the modal group index can also be calculated directly in the plane wave basis as

    \begin{equation}
        n_{\rm g,n} = \frac{ \omega_{\rm n}^2 + \frac{1}{2}\omega_{\rm n} \sum \Tilde{h}_{\rm n}^\dagger \left[ \mathbf{C}^\dagger \vep^{-1} \frac{d\vep}{d\omega} \vep^{-1} \mathbf{C} \right] \Tilde{h}_{\rm n} }{ \omega_{\rm n}\sum \Tilde{h}_{\rm n}^\dagger \frac{\partial \mathbf{M}}{\partial k} \Tilde{h}_{\rm n} } \label{eq:sup:pw_group_index}
    \end{equation}

\section{Modal Group Velocity Dispersion}\label{sec:sup:gvd}
The modal group velocity dispersion (GVD) is a quantity of central interest in dispersion engineering applications. 
Efficient differentiable calculation of modal GVD is thus valuable for dispersion optimizations. 
The GVD of the $\rm n$-th mode

    \begin{equation}
        \mathrm{GVD}_n = \frac{d n_{\rm g, n}}{d \omega} = \frac{d^2|k_{\rm n}|}{d \omega^2} 
    \end{equation}

can be approximated by finite difference in frequency

    \begin{equation}
        \mathrm{GVD}_n(\omega) \approx \frac{ n_{\rm g,n}(\omega + \delta\omega/2) - n_{\rm g,n}(\omega - \delta\omega/2) }{ \delta\omega }
    \end{equation}

however sensitivity analysis of GVDs calculated this way requires forward calculation and back-propagation through two independent eigenmode solutions. 
These redundant computations introduce excess numerical error due to uncorrelated finite errors in the independent eigenmode solutions..
The modal GVD can also be calculated by automatic differentiation using the adjoint method derived above for back-propagation through the eigenmode solution. 
Using this approach subsequent sensitivity analysis and optimization of GVDs and GVD-dependent quantities requires nested automatic differentiation which is error-prone and often infeasibly resource intensive for problems of this size and complexity.

We calculate this derivative without automatic differentiation -- it is more efficient to solve an adjoint equation from Eq.~\ref{eq:sup:pw_group_index} parameterized only in $\omega$. 
This method of GVD calculation allows subsequent application of automatic differentiation for sensitivity analysis of the GVD w.r.t geometric/material design parameters and is more efficient and accurate than alternative differentiable GVD calculation approaches based on finite differences in optical frequency.

We first calculate partial derivatives of the modal group index (Eq.~\ref{eq:sup:pw_group_index}) with respect to frequency $\omega$, the mode wave vector magnitude $k_{\rm n}$ and field $\Tilde{h}_{\rm n}$, the dielectric tensor $\vep$ and its first frequency derivative $\frac{d\vep}{d\omega}$

    \begin{equation}
        \frac{\partial n_{\rm g,n}}{\partial \omega} = \frac{\omega}{P} \label{eq:sup:EqPartialNgPartialOmega}
    \end{equation}

    \begin{equation}
        \frac{\partial n_{\rm g,n}}{\partial k_{\rm n}} =  \frac{\omega}{P} \sum \Tilde{h}_{\rm n}^\dagger \left[ 
            2 \mathbf{C}^\dagger \vep^{-1} \frac{d \vep}{d \omega} \vep^{-1}  \frac{d \mathbf{C}}{dk}
            -  n_{\rm g,n} \left( \frac{d \mathbf{C}^\dagger}{dk} \vep^{-1}  \frac{d \mathbf{C}}{dk} + \mathbf{C}^\dagger \vep^{-1}  \frac{d^2 \mathbf{C}}{dk^2} \right)
            \right] \Tilde{h}_{\rm n} \label{eq:sup:EqPartialNgPartialK}
    \end{equation}

    \begin{equation}
        \frac{\partial n_{\rm g,n}}{\partial \Tilde{h}_{\rm n}} = \frac{\omega}{P}\left( \frac{1}{2} \mathbf{C}^\dagger  \vep^{-1} \frac{d \vep}{d \omega} \vep^{-1} \mathbf{C} \Tilde{h}_{\rm n} + n_{\rm g,n} \mathbf{C}^\dagger \vep^{-1} \frac{d\mathbf{C}}{dk} \right) \Tilde{h}_{\rm n} + \mathrm{h.c.} \label{EqPartialNgPartialH}
    \end{equation}

    \begin{equation}
        \frac{\partial n_{\rm g,n}}{\partial \vep} = - \frac{ \omega }{ P } \vep^{-1} \left[ 
                \frac{d \vep}{d \omega} \vep^{-1} \mathbf{C} \Tilde{h}_{\rm n} \Tilde{h}_{\rm n}^\dagger \mathbf{C}^\dagger + 
                n_{\rm g,n} \vep^{-1}  \mathbf{C} \frac{d \mathbf{C}^\dagger}{dk} \vep^{-1}
            \right] \label{EqPartialNgPartialEps}
    \end{equation}

    \begin{equation}
        \frac{\partial n_{\rm g,n}}{\partial (d\vep / d\omega)} = \frac{ \omega }{ 2 P }  \vep^{-1} \mathbf{C} \Tilde{h}_{\rm n} \Tilde{h}_{\rm n}^\dagger \mathbf{C}^\dagger \vep^{-1} \label{EqPartialNgPartialDEpsDOmega}
    \end{equation}

where h.c. refers to Hermitian conjugate. We find $\frac{d^2 \mathbf{C}}{dk^2}$ by differentiating Eq.~\ref{EqBzDef} with respect to $k$

    \begin{equation}
        \frac{d^2 \mathbf{C}}{dk^2} = \mathbf{F} \begin{bmatrix} 0 & 1 & 0 \\ -1 & 0 & 0 \\ 0 & 0 & 0 \end{bmatrix} \mqty[ \frac{d\hat{m}}{dk} & \frac{d\hat{n}}{dk} ]    \label{Eqd_BzDef_dk}
    \end{equation}

where $\frac{d\hat{m}}{dk}$ and $\frac{d\hat{n}}{dk}$ follow from Eqs.~\ref{eqn:VjDef}, \ref{EqKtoN} \& \ref{EqKtoM} 

    \begin{equation}
        \frac{d\hat{n}}{d k} = 
        \frac{ \hat{a} \times \hat{z} - \left( \hat{n} \cdot (\hat{a} \times \hat{z}) \right) \hat{n}  }
             {              \left| \hat{a} \times (k_{\rm n} \hat{z} + \vec{g} )  \right|               }
    \end{equation}

    \begin{equation}
        \frac{d\hat{m}}{d k} = 
            \frac{ 
                \left[ \frac{d \hat{n}}{d k} \times (k \hat{z} + \vec{g} ) \right] + \left[ \hat{n} \times \hat{z} \right] - 
                \left( \hat{m} \cdot \left[ \frac{d \hat{n}}{d k} \times (k \hat{z} + \vec{g} ) + \hat{n} \times \hat{z} \right] \right) \hat{m}
            } 
            { \left| \hat{n} \times (k \hat{z} + \vec{g} )  \right| }
    \end{equation}
    
The modal GVD is the total derivative of the modal group index with respect to frequency. We calculate this in terms of the partial derivatives in Eqs.~\ref{eq:sup:EqPartialNgPartialOmega}-\ref{EqPartialNgPartialDEpsDOmega} and the first two frequency derivatives of the dielectric tensor $\frac{d\vep}{d\omega}$ and $\frac{d^2\vep}{(d\omega)^2}$

\begin{align}
    \begin{split}
    \mathrm{GVD}_{\rm n} = & \frac{\partial n_{\rm g,n}}{\partial \omega} 
            + \left\langle \frac{\partial n_{\rm g,n}}{\partial \Tilde{h}_{\rm n}}, \frac{\partial \Tilde{h}_{\rm n}}{\partial \omega} \right\rangle
            + \frac{\partial n_{\rm g,n}}{\partial k_{\rm n}} \frac{\partial k_{\rm n}}{\partial \omega}  \\
            + &\left\langle \left[ \frac{\partial n_{\rm g,n}}{\partial \vep} +
                \frac{\partial n_{\rm g,n}}{\partial k_{\rm n}} \frac{\partial k_{\rm n}}{\partial \vep} +
                \frac{\partial n_{\rm g,n}}{\partial \Tilde{h}_{\rm n}} \frac{\partial \Tilde{h}_{\rm n}}{\partial \vep}
            \right], \frac{d \vep}{d \omega} \right\rangle
            + \left\langle \frac{\partial n_{\rm g,n}}{\partial (d\vep / d\omega)}, \frac{d^2\vep}{(d\omega)^2} \right\rangle \label{EqGVDfromPartials}
    \end{split}
\end{align}

where $\frac{\partial n_{\rm g,n}}{\partial \Tilde{h}_{\rm n}} \frac{\partial \Tilde{h}_{\rm n}}{\partial \omega_{\rm n}} + \frac{\partial n_{\rm g,n}}{\partial k_{\rm n}} \frac{\partial k_{\rm n}}{\partial \omega_{\rm n}}$ and $\frac{\partial n_{\rm g,n}}{\partial \Tilde{h}_{\rm n}} \frac{\partial \Tilde{h}_{\rm n}}{\partial \vep} + \frac{\partial n_{\rm g,n}}{\partial k_{\rm n}} \frac{\partial k_{\rm n}}{\partial \vep}$ are calculated from $ \frac{\partial n_{\rm g,n}}{\partial k} $ (Eq.~\ref{eq:sup:EqPartialNgPartialK}) and $ \frac{\partial n_{\rm g,n}}{\partial \Tilde{h}_{\rm n}}$ (Eq.~\ref{EqPartialNgPartialH}) using the adjoint method described in Sections~\ref{SecLinearEigenproblemAdj}~\&~\ref{sec:sup:NonlinearEigenproblemAdj}. 
Note that the last two terms in Eq.~\ref{EqGVDfromPartials} imply sums over Frobenius inner products between $3\times3$ matrices at each spatial grid point.

\begin{figure}[h!]
    \centering{
    \phantomsubcaption\label{fig:sup:shg_opt:cross_section}
    \phantomsubcaption\label{fig:sup:shg_opt:w_top}
    \phantomsubcaption\label{fig:sup:shg_opt:t_core}
    \phantomsubcaption\label{fig:sup:shg_opt:etch_frac}
    \phantomsubcaption\label{fig:sup:shg_opt:cost_fn}
    \phantomsubcaption\label{fig:sup:shg_opt:poling_period}
    \phantomsubcaption\label{fig:sup:shg_opt:shg_trnsfr}
    }
    \includegraphics[width=\textwidth]{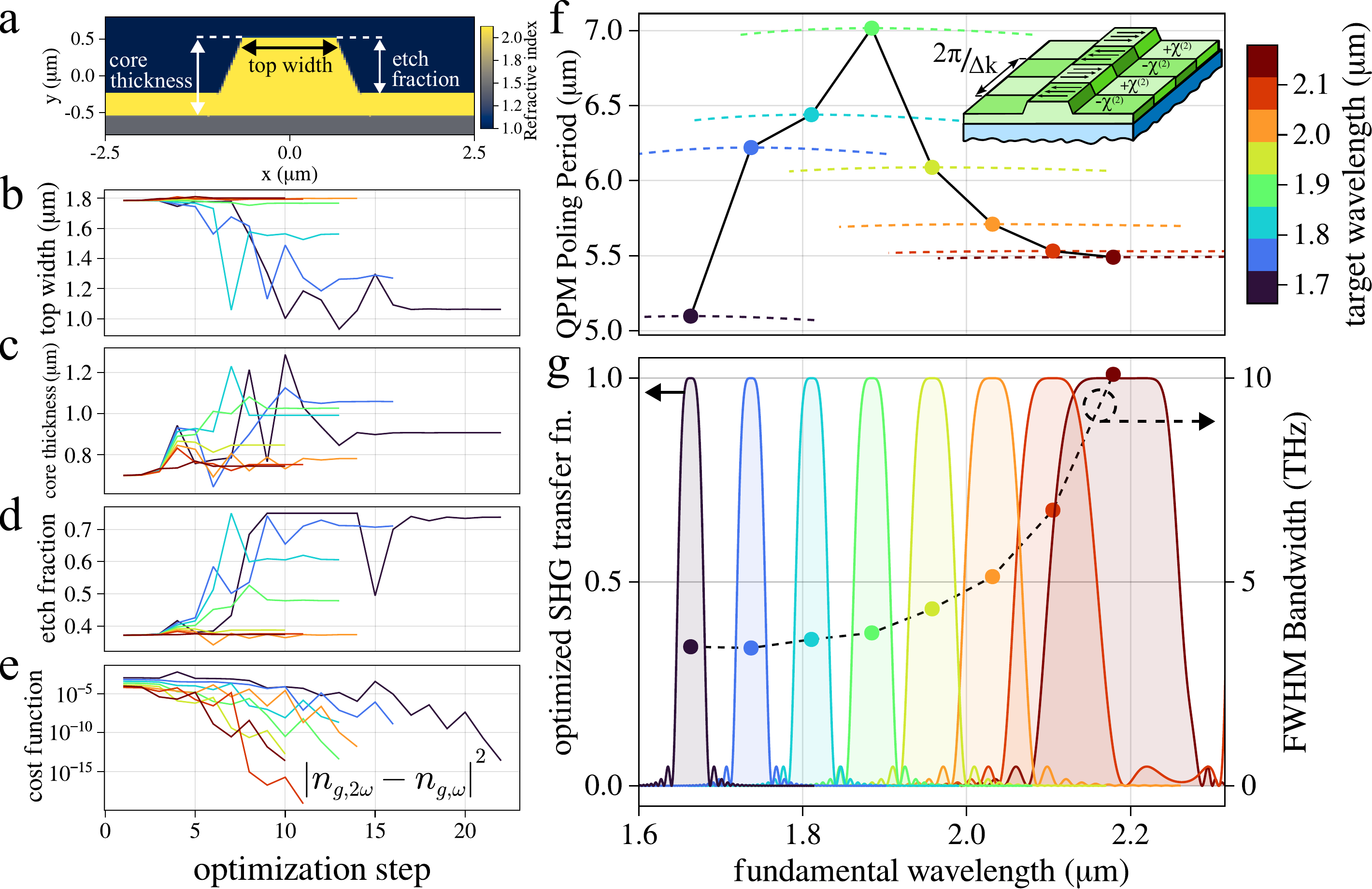}
    \caption{
    Optimization of partially-etched LiNbO$_3$ rib waveguide geometry for maximized second harmonic generation (SHG) quasi-phase-matching (QPM) bandwidth with quasi-TE$_{\rm 0,0}$ modes.
    \subref{fig:sup:shg_opt:cross_section} Waveguide cross section showing the three geometry parameters used for optimization: rib top width, full (unetched) thickness of the LiNbO$_3$ core, and partial etch depth in the rib cladding expressed as a fraction of the full LiNbO$_3$ layer thickness.  
    \subref{fig:sup:shg_opt:w_top}-\subref{fig:sup:shg_opt:cost_fn} Variation of optimized geometry parameters and the cost function during optimizations of SHG QPM bandwidth at several target wavelengths. The cost function is the square magnitude of the modal group velocity mismatch (GVM) between fundamental and second harmonic waves at a target fundamental wavelength. 
    \subref{fig:sup:shg_opt:poling_period} Poling period for quasi-phase-matched SHG at the target (dots) and nearby wavelengths (dashed lines) in optimized waveguide geometries for each target wavelength. 
    \subref{fig:sup:shg_opt:shg_trnsfr} Relative SHG transfer functions (filled traces) and FWHM phase matching bandwidths (dots) of optimized waveguide geometries for each target wavelength. 
    }
    \label{fig:sup:shg_opt}
\end{figure}

\section{SHG Bandwidth Optimization}\label{sec:sup:shg_opt}

Figure~\ref{fig:sup:shg_opt} shows eight independent optimizations of LiNbO$_3$ rib waveguide geometry for maximized SHG QPM bandwidth for fundamental vacuum wavelengths between 1.7-2.2~$\mu$m.
In each optimization the square magnitude of the group velocity mismatch (GVM) between fundamental and second harmonic waves at a different target fundamental wavelength was used as the cost function.
All optimizations were initiated from the same nominal waveguide geometry parameters.
Figure~\ref{fig:sup:shg_opt:cost_fn} demonstrates significant improvement over 10-30 steps for every target wavelength.
Periodic poling, depicted in the inset, reverses the sign of the second order nonlinear polarizability ($\chi^{(2)}$) to compensate for mismatched phase velocities of fundamental and second harmonic waves.
The first frequency derivative of the poling period required for SHG QPM is proportional to GVM, and is thus minimized near the target wavelength in these optimizations.
This is shown in Figure~\ref{fig:sup:shg_opt:poling_period}. 
Poling does not affect linear optical dispersion and thus the poling period can be chosen freely after modal dispersion optimization.

\begin{figure}[h!]
    \centering{
    \phantomsubcaption\label{fig:sup:gvd_opt:w_top}
    \phantomsubcaption\label{fig:sup:gvd_opt:t_core}
    \phantomsubcaption\label{fig:sup:gvd_opt:etch_frac}
    \phantomsubcaption\label{fig:sup:gvd_opt:cost_fn}
    \phantomsubcaption\label{fig:sup:gvd_opt:gvd_spectra}
    }
    \includegraphics[width=\textwidth]{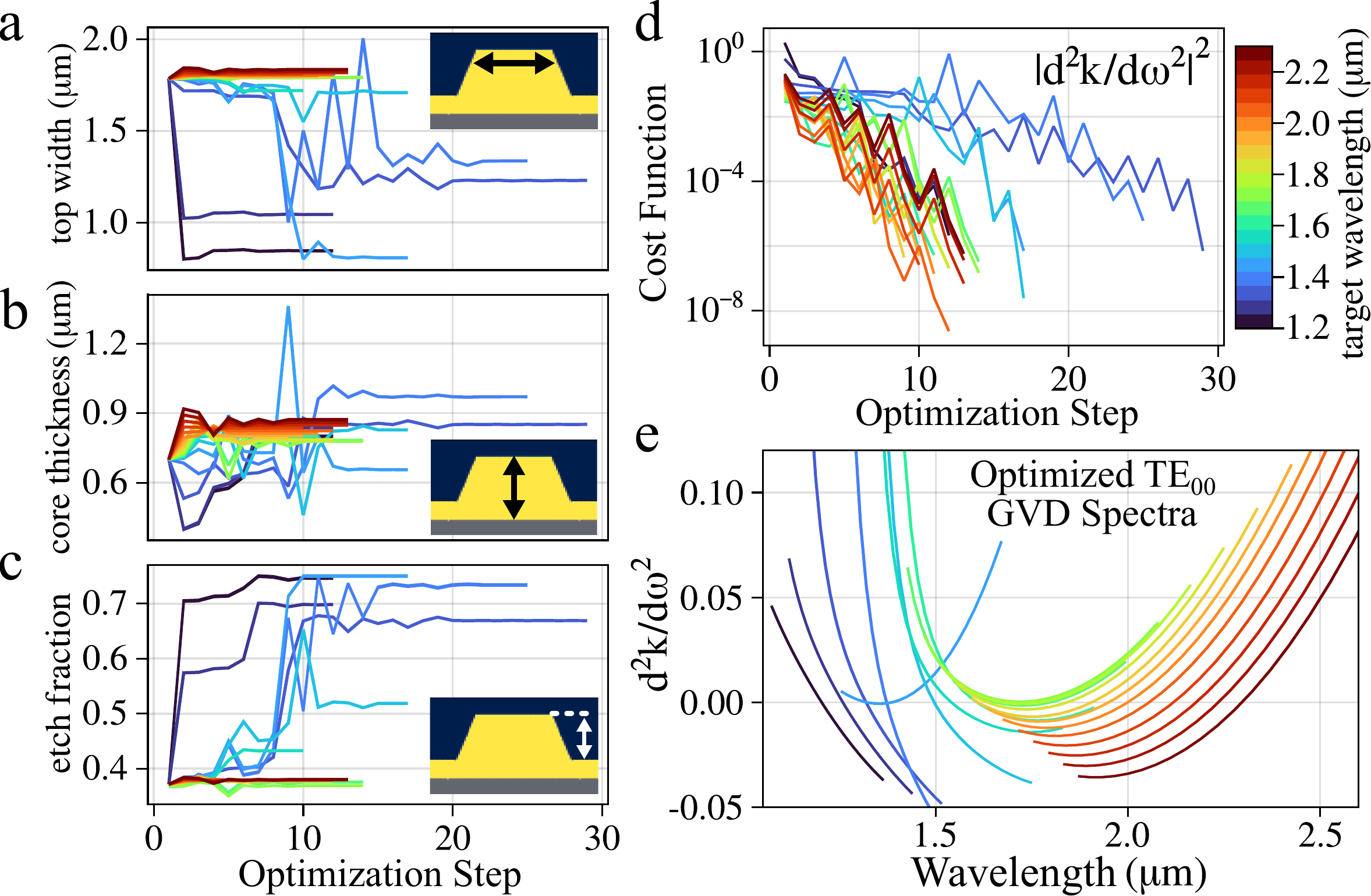}
    \caption{
    Optimization of partially-etched LiNbO$_3$ rib waveguide geometry for zero quasi-TE$_{\rm 0,0}$ modal group velocity dispersion at target wavelengths between 1.2-2.3~$\mu$m.
    \subref{fig:sup:gvd_opt:w_top}-\subref{fig:sup:gvd_opt:etch_frac} Variation of optimized geometry parameters (width, thickness and etch depth) during optimizations modal GVD at each target wavelength. 
    \subref{fig:sup:gvd_opt:cost_fn} Evolution of the cost function during optimization at each target wavelength. The cost function in each optimization was the square magnitude of the modal group velocity dispersion (GVD) at the target wavelength.
    \subref{fig:sup:gvd_opt:gvd_spectra} GVD spectra of each optimized waveguide geometry. In each case modal GVD at the target wavelength is approximately zero. 
    }
    \label{fig:sup:gvd_opt}
\end{figure}

\section{GVD Optimization}\label{sec:sup:gvd_opt}

Figure~\ref{fig:sup:gvd_opt} shows twenty independent optimizations of LiNbO$_3$ rib waveguide geometry for maximized SHG QPM bandwidth for fundamental vacuum wavelengths between 1.2-2.3~$\mu$m.
In each optimization the square magnitude of the TE$_{0,0}$ modal group velocity mismatch (GVD) at a different target wavelength was used as the cost function.
These calculations were performed using normalized units (vacuum light speed $c_0$=1) so that optical frequencies and modal wave vectors have units of $\mu$m$^{-1}$ and the GVD is unitless.
All optimizations were initiated from the same nominal waveguide geometry parameters.
Figure~\ref{fig:sup:shg_opt:cost_fn} shows a reduction in the square magnitude of the GVD by 2-6 orders of magnitude over 12-30 steps for every target wavelength.

\bibliography{references}